\documentclass[useAMS,usenatbib]{mn2e}
\usepackage{times}
\usepackage{bigints}
\usepackage[pdftex]{graphicx}
\usepackage{amsmath, mathtools}
\usepackage{amssymb}
\usepackage{natbib}
\usepackage{hyperref} 
\usepackage{newlfont,verbatim,enumerate,array}
\usepackage[usenames,dvipsnames,svgnames,table]{xcolor}
\RequirePackage{rotating}
\newcommand\vect[1]{\ensuremath{\mathbf{#1}}}

\newcommand\velocity{\ensuremath{\text{km~s}^{-1}}}
\newcommand\patspeed{\ensuremath{\text{km~s}^{-1}\text{~kpc}^{-1}}}
\newcommand\spiralampl{\ensuremath{\text{km}^2\text{~s}^{-2}\text{~kpc}^{-1}}}
\newcommand\amuse{\textsc{Amuse}}
\newcommand\bridge{\textsc{Bridge}}

\newcommand\gaia{\textit{Gaia}}
\newcommand\galah{\textit{GALAH}}
\newcommand\SF{\ensuremath{f_\mathrm{sib}}}
\newcommand\kms{\ensuremath{\text{km~s}^{-1}}}
\newcommand\maspyr{\ensuremath{\text{mas~yr}^{-1}}}

\newcommand\pbirth{\ensuremath{p(\vect{x}_\mathrm{b}, \vect{v}_\mathrm{b})}}

\newcommand\olrbar{\ensuremath{\text{OLR}_\mathrm{bar}}}
\newcommand\corotation{\text{CR}}
\newcommand\crsp{\ensuremath{\text{CR}_\mathrm{sp}}}

\newcommand\mc{\ensuremath{M_\mathrm{c}}}
\newcommand\rc{\ensuremath{R_\mathrm{c}}}
\newcommand\binaryf{primordial binary fraction}
\newcommand\modela{model \textbf{a}}
\newcommand\modelb{model \textbf{b}}
\newcommand\modelc{model \textbf{c}}
\newcommand\modeld{model \textbf{d}}
\newcommand\centermass{\ensuremath{ \left(\vect{x}_\mathrm{cm}, \vect{v}_\mathrm{cm} \right)}}

\title[The evolution of the Sun's birth cluster and the search for the solar siblings]{The evolution of the Sun's birth cluster and the search for the solar siblings with \gaia}
\author[C.A. Mart\'inez-Barbosa et al.]{C.A. Mart\'inez-Barbosa,$^{1}$\thanks{E-mail:
cmartinez@strw.leidenuniv.nl} A.G.A Brown,$^{1}$\thanks{E-mail:
brown@strw.leidenuniv.nl} T. Boekholt,$^{1}$ S. Portegies Zwart,$^{1}$ E. Antiche$^{2}$\vspace{0.3 cm}\\ {\normalfont\LARGE and T. Antoja$^{3}$}\\
$^{1}$ Leiden Observatory, Leiden University, P.B. 9513, Leiden 2300 RA, the Netherlands \\
$^{2}$ Departament d' Astronomia i Meteorologia, Institut de Ci\`encies del Cosmos, Universitat de Barcelona, IEEC, Mart\'i Franqu\`es 1, E-08028 Barcelona, Spain\\
$^3$ Scientific Support Office, Directorate of Science and Robotic Exploration, European Space Research and Technology Centre (ESA/ESTEC), Keplerlaan 1, \\ NL-2201 AZ Noordwijk, the Netherlands}

\begin{document}

\date{Accepted XXXXXXXXXXX. Received XXXXXXXXXXXX; in original form XXXXXXXXXXXXX}

\pagerange{\pageref{firstpage}--\pageref{lastpage}} \pubyear{2002}

\maketitle

\label{firstpage}

\begin{abstract}
We use self-consistent numerical simulations of the evolution and disruption of the Sun's birth
cluster in the Milky Way potential to investigate the present-day phase space distribution of
the sun's siblings. The simulations include the gravitational $N$-body forces within the cluster
and the effects of stellar evolution on the cluster population. In addition the gravitational
forces due to the Milky Way potential are accounted for in a self-consistent manner. Our aim is to
understand how the astrometric and radial velocity data from the \gaia\ mission can be used to
pre-select solar sibling candidates. We vary the initial conditions of the Sun's birth cluster, as
well as the parameters of the Galactic potential. In particular, we use different configurations
and strengths of the bar and spiral arms. We show that the disruption time-scales of the cluster
are insensitive to the details of the non-axisymmetric components of the Milky Way model and we
make predictions, averaged over the different simulated possibilities, about the number of solar
siblings that should appear in surveys such as \gaia\ or \galah. We find a large variety of
present-day phase space distributions of solar siblings, which depend on the cluster initial
conditions and the Milky Way model parameters. We show that nevertheless robust predictions can be
made about the location of the solar siblings in the space of parallaxes ($\varpi$), proper
motions ($\mu$) and radial velocities ($V_\mathrm{r}$). By calculating the ratio of the number of
simulated solar siblings to that of the number of stars in a model Galactic disk, we find that
this ratio is above $0.5$ in the region given by: $ \varpi \geq 5$~mas, $4 \leq \mu \leq 6$~mas
yr$^{-1}$, and $-2\leq V_\mathrm{r} \leq 0$~km s$^{-1}$. Selecting stars from this region should
increase the probability of success in identifying solar siblings through follow up observations.
However the proposed pre-selection criterion is sensitive to our assumptions, in particular about
the Galactic potential. Using a more realistic potential (e.g., including transient spiral
structure and molecular clouds) would make the pre-selection of solar sibling candidates based on
astrometric and radial velocity data very inefficient. This reinforces the need for large scale
surveys to determine precise astrophysical properties of stars, in particular their ages and
chemical abundances, if we want to identify the solar family.
\end{abstract}


\section{Introduction}\label{Sect:introd}

Since most of the stars are born in star clusters \citep{lada}, these systems are considered the
building blocks of galaxies. In the Milky Way star clusters located in the Galactic halo (Globular
clusters) populate the Galactic disk through mergers \citep{merging}. On the other hand star
clusters formed in the Galactic disk (open clusters) supply new stars to the disk of the Galaxy
through several processes, such as shocks from encounters with spiral arms and Giant Molecular
Clouds \citep{gieles06, gieles07}.

The dynamical evolution of star clusters involves several physical mechanisms. At earlier stages of
their evolution, star clusters lose mass mainly due to stellar evolution and two-body relaxation
processes, which in turn, enlarge the size of star clusters \citep{takashi00, BM03, madrid12}.  This
evolutionary stage is called the expansion phase \citep{gieles11}, which takes about $40$\% of the
star cluster's lifetime. Once star clusters overcome the expansion phase, the effects of the
external tidal field of the Galaxy become important, depending on their location with respect to the
Galactic centre. This stage is called the evaporation phase \citep{gieles11} and it is characterized
by the gradual dissolution of star clusters in the Galaxy.

The dissolution rate of star clusters depends on their Galactocentric distance \citep{madrid12},
orbit \citep{BM03}, orbital inclination \citep{webb14} and on Galaxy properties such as the mass and
size of the Galactic disk \citep{madrid14}. Additionally, open clusters in the Milky Way are also
dissolved due to non axisymmetric perturbations such as bars \citep{cluster_bar}, spiral arms
\citep{gieles07} and giant molecular clouds \citep{gieles06, lammers06}. The strongest tidal
stripping occurs at times when open clusters cross regions of high density gas, for instance, during
spiral arms passages \citep{gieles07,kruijssen} or during collisions with giant molecular clouds
\citep{gieles06}.  Open clusters can also radially migrate over distances of up to $1$~kpc in a
short time scale ($\sim 100$~Myr) when the Galactic spiral structure is transient \citep{fujii12}.
This radial migration process can also be efficient in the absence of transient structure if the
resonances due the bar and spiral structure overlap \citep{minchev10}. Radial migration affects the
orbits of open clusters in the Galaxy, increasing or decreasing their perigalacticon distance, which
in turn influences their dissolution times \citep[see e.g.][]{jilkova12}.

The high eccentricities and inclinations observed in the Edgeworth-Kuiper belt objects together with
the discovery of decay products of $^{60}$Fe and other radioactive elements in the meteorite fossil
record, suggest that the Sun was born in an open cluster $4.6$~Gyr ago \citep[and references
therein]{portegies09}. Identifying the stars that were formed together with the Sun (the solar
siblings) would enable the determination of the Galactic birth radius of the Sun as well as further
constrain the properties of its birth cluster \citep{BH10,adams}. The birth radius affects the
evolution of the solar system, and in particular the Oort cloud, which is sensitive to the Galactic
environment the Sun passes through along its orbit \citep[e.g.][]{spzjilkova15}.

The Sun's birth cluster will undergo all the disruptive processes described above
and thus dissolve, leading to the spreading out of its stars over the Galactic disk. The subsequent
distribution of the solar siblings was studied by \cite{portegies09}, who evolved the Sun's birth
cluster in an axisymmetric model for the Galactic potential and concluded that tens of solar
siblings might still be present within a distance of $100$~pc from the Sun. Several attempts have
since been made to find solar siblings \citep[e.g.][]{brown10, bobylev11, liu15}; however, only four
plausible candidates have been identified so far \citep{batista12, batista14, ramirez14}. This small
number of observed solar siblings might be a consequence of the lack of accurate predictions of the
present-day phase space distribution of solar siblings together with insufficiently accurate stellar
kinematic data.

\cite{brown10} used test particle simulations to predict the current distribution of solar siblings
in the Milky Way. They concluded that stars with parallaxes ($\varpi$) $\geq 10$~mas and proper
motions ($\mu$) $\leq 6.5$~mas yr$^{-1}$, should be considered solar sibling candidates.  Their
conclusions were criticised by \cite{mishurov11} who pointed out that in more realistic Galactic
potentials the solar siblings are expected to be much more spread out over the Galactic disk. For
small birth clusters (few thousand stars with a total mass of the order of 1000~$M_\odot$) such as
employed by \cite{brown10} and \cite{portegies09}, \cite{mishurov11} predict that practically no
solar siblings will currently be located within 100~pc from the sun.  However, for larger birth
clusters \citep[of order $10^4$ stars, in line with predictions from e.g.][]{dukes} one can still
expect to find a good number of siblings presently orbiting the Galaxy within 100~pc from the Sun.

Ongoing surveys of our galaxy, in particular the {\gaia} mission \citep{gaia} and the {\galah}
survey \citep{galah}, will provide large samples of stars with accurately determined distances,
space motions, and chemical abundance patterns, thus enabling a much improved search for the sun's
siblings. In this paper we investigate the potential of the {\gaia} astrometric and radial velocity
data to narrow down the selection of candidate solar siblings for which detailed chemical abundance
studies should be undertaken in order to identify the true siblings. Our investigation is done by
performing simulations of the evolution and disruption of the Sun's birth cluster in a realistic (although
static) Galactic potential, including the bar and spiral arms. The aim is to predict the present-day
phase space distribution of the siblings and simulate the astrometric and radial velocity data
collected by {\gaia}. We include the internal $N-$body processes in the cluster to account for the
disruption time scale. We use a full stellar mass spectrum and a parametrized stellar evolution code
to make accurate predictions of how the solar siblings are observed by {\gaia}. To this end we also
account for the effects of extinction and reddening.

The rest of this paper is organized as follows. In Sect.\ \ref{sect:sim} we describe the
simulations. In Sect.\ \ref{sect:ev_sbc} we explore the evolution and disruption of the Sun's birth
cluster due to the bar and spiral arms of the Galaxy. In Sect.\ \ref{sect:distrib} we present the
current phase-space distribution of solar siblings obtained from the simulations.  In Sect.\
\ref{sect:kss} we make use of the simulated positions and motions of the solar siblings to
investigate the robustness of the selection criterion proposed by \cite{brown10} to the
uncertainties in the present-day phase space distribution of the solar siblings. An updated set of
selection criteria based on parallax, proper motion and radial velocity information is presented. In
Sect.\ \ref{sect:discuss} we use these criteria to examine stars that were previously suggested as
solar siblings candidates and further discuss our results. In Sect.\ \ref{sect:concl} we summarize.

\section{Simulation set-up}
\label{sect:sim}

The goals of the simulations of the Sun's birth cluster are to predict the present-day phase space
distribution of the solar siblings and how these are expected to appear in the {\gaia} catalogue. In
particular we wish to account for the uncertainties in the initial conditions of the birth cluster
and the parameters of the Milky Way potential. The predictions of the {\gaia} observations require
the use of a realistic mass spectrum for the siblings, and accounting for stellar evolution and
extinction and interstellar reddening effects. We thus employ the following elements in the
simulations:
\begin{description}
  \item[\textit{Galactic model}] The Milky Way potential is described by an analytic model
    containing a disk, bulge and halo, as well as a bar and spiral arms. The parameters of the bar
    and spiral arms are varied in the simulations to account for uncertainties in their strengths
    and pattern speeds (Sect.\ \ref{sect:Gmodel}).
  \item[\textit{Cluster model}] The Sun's birth cluster is modelled with a  mass spectrum for the
    stars and we account for the gravitational $N$-body effects within the cluster as well as the
    effect of the Galaxy's gravitational field on the cluster stars. The use of  $N$-body models for
    the birth cluster is motivated by the desire to account for the disruption time of the cluster
    which can be a substantial fraction of the lifetime of the Sun (Sect.\ \ref{sect:bcluster}).
  \item[\textit{Stellar evolution}] Predicting the observations of the Sun's birth cluster by
    {\gaia} requires that we account for the mass-dependent evolution of the solar siblings, in
    order to obtain the correct present-day apparent magnitudes and colours which are used to
    predict which stars end up in the {\gaia} catalogue. This prediction also requires us to account
    for interstellar extinction and reddening for which we employ a Galactic extinction model
    (Sects.\ \ref{sect:simulations}, \ref{sect:kss}).
\end{description}
These elements are described in more detail in the subsequent subsections.

\begin{figure*}
  \centering
  \includegraphics[width= 18cm, height= 6cm]{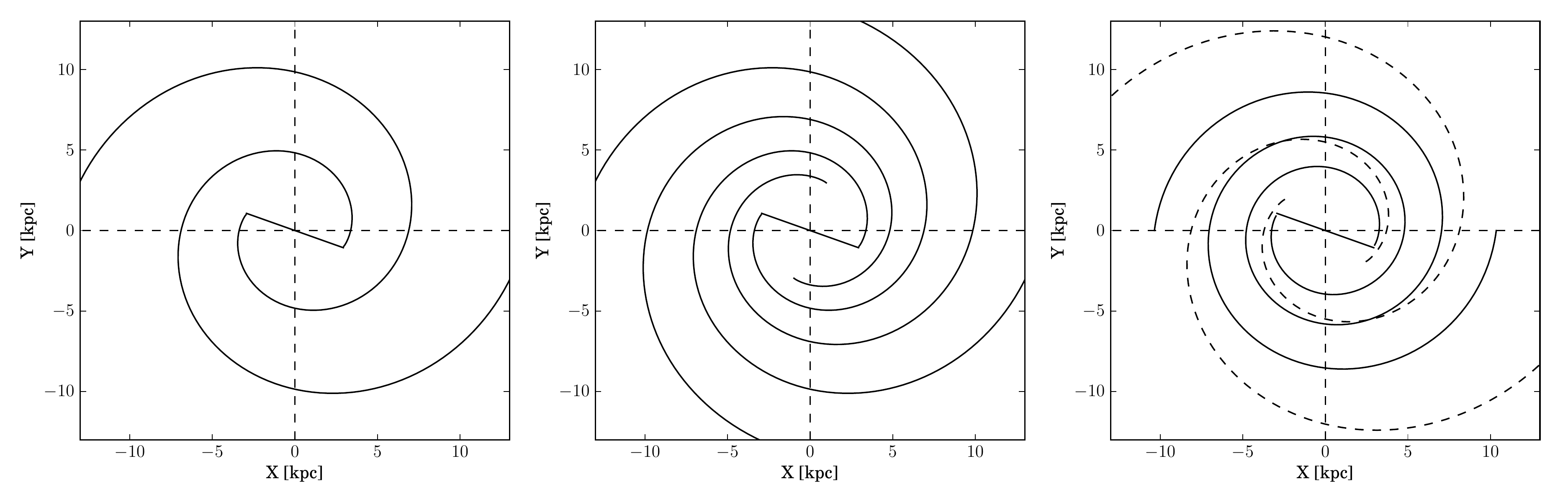}
  \caption{Configurations of the Galactic potential at the present time. \textit{Left:} Galaxy with
  two spiral arms. \textit{Middle:} Galaxy with four spiral arms. \textit{Right:} $(2+2)$ composite
  model. \label{fig:potentials} }
\end{figure*}

\subsection{Galactic model}
\label{sect:Gmodel}

We use an analytical potential to model the Milky Way. This potential contains two parts: an
axisymmetric component, which corresponds to a bulge, disk and a dark matter halo, and a
non-axisymmetric component which includes a central bar and spiral arms. Bellow we explain these
components in more detail.

\begin{table}
  \caption{Parameters of the Milky Way model potential.}
  \label{tab:galparams_p2}
  \begin{tabular}{ll} \hline
    \multicolumn{2}{c}{\textit{ Axisymmetric component}} \\
    Mass of the bulge ($M_\mathrm{b}$) & $1.41\times 10^{10}$ $M_{\odot}$ \\
    Scale length bulge ($b_\mathrm{1}$) & $0.38$ kpc\\
    Disk mass ($M_\mathrm{d}$) & $8.56\times10^{10}$ $M_{\odot}$\\
    Scale length disk 1 ($a_\mathrm{2}$) & $5.31$ kpc\\
    Scale length disk 2 ($b_\mathrm{2}$) & $0.25$ kpc\\
    Halo mass ($M_\mathrm{h}$) & $1.07\times 10^{11} $ $M_{\odot}$\\
    Scale length halo ($a_\mathrm{3}$) & 12 kpc\\
    \multicolumn{2}{c}{\textit{Central Bar}} \\
    Pattern speed ($\Omega_\mathrm{bar}$) & $40$--$70$ km~s$^{-1}$~kpc$^{-1}$\\
    Semi-major axis ($a$) & $3.12$ kpc \\
    Axis ratio ($b/a$) & $0.37$ \\
    Mass ($M_\mathrm{bar}$) & $9.8\times10^{9}$--$1.4\times10^{10}$ $M_{\odot}$\\
    Present-day orientation & $20^\circ$ \\
    Initial orientation & $1^\circ-167^\circ$\\ 
    \multicolumn{2}{c}{\textit{ Spiral arms}} \\
    Pattern speed ($\Omega_\mathrm{sp}$) & $15$--$30$ km~s$^{-1}$~kpc$^{-1}$\\
    Locus beginning ($R_\mathrm{{sp}}$) & $3.12$ kpc\\
    Number of spiral arms ($m$) & $2$, $4$\\
    Spiral amplitude ($A_\mathrm{sp}$) & $650$--$1100$ \spiralampl \\
    Pitch angle ($i$) & $ 12.8^\circ$ \\
    Scale length ($R_\mathrm{{\Sigma}}$) & $2.5$ kpc\\
    Present-day orientation & $20^\circ$ \\ 
    Initial orientation & $103^\circ-173^\circ$\\ \hline
      \end{tabular}
\end{table}

\begin{table}
  \caption{Parameters of the composite Galaxy model potential.}
  \label{tab:composite_p2}
  \begin{tabular}{ll} \hline
        \multicolumn{2}{c}{\textit{Main spiral structure} }\\
     Pattern speed ($\Omega_{\mathrm{sp}_1}$) & $26$ \patspeed\\
     Amplitude ($A_{\mathrm{sp}_1}$) & $650$--$1300$ \spiralampl \\
     Pitch angle ($i_1$) & $-7^\circ$ \\
     Present-day orientation & $20^\circ$ \vspace{1mm}\\ 	
     Initial orientation & $171^\circ$ \\	
      \multicolumn{2}{c}{\textit{Secondary spiral structure}}\\
     Pattern speed ($\Omega_{\mathrm{sp}_2}$) & $15.8$ \patspeed\\
     Amplitude ($A_{\mathrm{sp}_2}$) & $0.8 A_{\mathrm{sp}_1}$ \\
     Pitch angle ($i_2$) & $-14^\circ$ \\
     Present-day orientation & $220^\circ$ \\
      Initial orientation & $158^\circ$ \\	
      \multicolumn{2}{c}{\textit{Bar} }\\
    Pattern speed ($\Omega_\mathrm{bar}$) & $40$ \patspeed\\
    Semi-major axis ($a$) & $3.12$ kpc \\
    Axis ratio ($b/a$) & $0.37$ \\
    Mass ($M_\mathrm{bar}$) & $9.8\times10^{9}$ $M_{\odot}$\\
    Strength of the bar ($\epsilon_\mathrm{b}$) & $0.3$ \\
    Present-day orientation & $20^\circ$ \\ 
     Initial orientation & $1^\circ$\\ \hline
 \end{tabular}
\end{table}

\paragraph*{Axisymmetric component}

We use the potential of \cite{allen} to model the axisymmetric component of the Galaxy. In this
approach, the bulge is modelled with a Plummer \citep{plummer} potential; the disk is modelled with
a Miyamoto-Nagai \citep{miyamoto} potential and the dark matter halo with a logarithmic potential.
The parameters used to model the axisymmetric component of the Galaxy are listed in table
\ref{tab:galparams_p2}.

The model introduced by \cite{allen} predicts a rotational velocity of $220$~\velocity\ at the solar
radius, which does not match with the recent observational estimates \cite[see
e.g][]{mcmillan1,reid14}. However, \cite{jilkova12} did not find substantial variations in the
orbits of open clusters when using different models of the axisymmetric structure of the Galaxy.
Therefore, we do not expect that the evolution of the Sun's birth cluster and the present-day
distribution of solar siblings will be affected due to the choice of the axisymmetric potential
model.

\paragraph*{The Galactic bar}

The central bar is modelled with a Ferrers potential \citep{ferrers} which describes the potential
associated to an elliptical distribution of mass. In an inertial frame located at the Galactic
centre, the bar rotates with a constant pattern speed of $40$--$70$~\patspeed\ \citep{martinezb14}.
This range of angular velocities places the Outer Lindblad resonance of the bar (\olrbar) at
$10$--$5$~kpc from the Galactic centre. In the same inertial frame,  the present-day orientation of
the bar with respect to the negative $x$-axis is $20^\circ$ \citep[and references
therein]{pichardo04, pichardo1, merce2}. In the left panel of Fig.\ \ref{fig:potentials} we show the
present-day orientation of the Galactic bar. In Table \ref{tab:galparams_p2} we show the parameters
used in this study. For further details on the choice of the bar parameters, we refer the reader to
\cite{martinezb14}.

\paragraph*{The spiral arms} We model the spiral arms as periodic perturbations of the axisymmetric
potential \citep[tight winding approximation,][]{lin}.  The spiral arms rotate with a constant
pattern speed of $15$--$30$~\patspeed\ \citep{martinezb14}. This range of values places the
co-rotation resonance of these structures (\crsp) at $14$--$7$~kpc from the Galactic centre.  We
assume that the Galaxy has two or four non-transient spiral arms with the same amplitude. A
schematic picture of the present-day configuration of the spiral arms is shown in the left and
middle panels of Fig.\ \ref{fig:potentials}. The parameters of the spiral arms used in this study
are listed in Table \ref{tab:galparams_p2}.  For further details on the choice of these parameters,
we refer the reader to \cite{martinezb14}.

\paragraph*{Initial orientation of the bar and spiral arms}
The orientation of the bar and spiral arms at the beginning of the simulations (i.e $4.6$~Gyr ago)
are defined through the following equations:
\begin{align}
\label{eq:orientation}
\varphi_\mathrm{b} &= \varphi_\mathrm{b}(0) - \Omega_\mathrm{bar}\mathrm{t}\,, \nonumber \\
\varphi_\mathrm{s} &= \varphi_\mathrm{s}(0) - \Omega_\mathrm{sp}\mathrm{t}\,.
\end{align}
Here $\varphi_\mathrm{b}(0)$ is the present-day orientation of the bar. We assume that the spiral
arms start at the tips of the bar, i.e. $\varphi_\mathrm{s}(0)= \varphi_\mathrm{b}(0)$ (see Fig.
\ref{fig:potentials}). The time, $t= 4.6$~Gyr corresponds to the age of the Sun \citep{buonano}.
The initial orientations of the bar and spiral arms are listed in Table \ref{tab:galparams_p2}.

\paragraph*{Multiple spiral patterns}
We also consider a more realistic Galaxy model with multiple spiral patterns, as suggested by
\cite{lepine11}. In this model, often called the $(2+2)$ composite model, two spiral arms have a
smaller amplitude and pattern speed than the main structure, which is also composed of two spiral
arms.  A schematic picture of the composite model is shown in the right panel of Fig.
\ref{fig:potentials}. We use the parameters of the composite model suggested by \cite{mishurov11}
and \cite{lepine11}. These values are listed in Table \ref{tab:composite_p2}. Here, $
A_{\mathrm{sp}_1}$ corresponds to a strength of $0.06$; that is, the main spiral structure has 6\%
the strength of the axisymmetric potential. Additionally, the value of $\Omega_{\mathrm{sp}_1}$
places the co-rotation resonance (\corotation) of the main spiral structure at the solar radius.
The value of $\Omega_{\mathrm{sp}_2}$ on the other hand, places the \corotation\ of the secondary
spiral structure at $13.6$~kpc. The orientation of the spiral arms at the beginning of the
simulation is set according to Eq.\ \ref{eq:orientation}, where $\varphi_{0s_1}= 20^\circ$ and
$\varphi_{0s_2}= 220^\circ$ are the initial phases of the main and secondary spiral structures
respectively. In the composite model we also fixed the parameters of the bar. The corresponding
values are listed in Table \ref{tab:composite_p2}.

\subsection{The Sun's birth cluster}
\label{sect:bcluster}

\subsubsection{Initial conditions}
We model the Sun's birth cluster with a spherical density distribution corresponding to a Plummer
potential \citep{plummer}. We also assume that the primordial gas was already expelled from the
cluster when it starts moving in the Galaxy. The initial mass (\mc) and radius (\rc) of the Sun's
birth cluster were set according to \cite{portegies09}, who suggested that the Sun was probably born
in a cluster with $\mc= 500$--$3000$~$M_{\odot}$ and $\rc= 0.5$--$3$~pc. In table
\ref{tab:ic_clusters} we show the initial mass and radius of the Sun's birth cluster used in the
simulations. From this table we note that the number of stars belonging to the Sun's birth cluster
($N$) is around $10^{2}$--$10^3$ in accordance with previous studies \citep[see e.g.][]{adams01,
adams}. In table \ref{tab:ic_clusters} we also show the initial velocity dispersion of the Sun's
birth cluster ($\sigma_\mathrm{v}$). This quantity can be computed by means of the virial theorem.
As can be observed, for the initial mass and radius adopted, $\sigma_\mathrm{v}$ is between $1.4$
and $2.9$ \velocity.

We used a Kroupa initial mass-function (IMF) \citep{kroupa} to model the mass distribution of the
Sun's birth cluster. The minimum and maximum masses used are $0.08$~$M_\odot$ and $100$~$M_\odot$
respectively. In this regime the IMF is a two-power law function described by the relation:

\begin{equation}
  \psi(m)=
  \begin{cases}
    A_1m^{-1.3} & 0.08< m \leq 0.5\text{~}M_\odot, \\
    A_2m^{-2.3} &  m > 0.5\text{~}M_\odot.
  \end{cases}\
\end{equation}
Here $A_1$ and $A_2$ are normalization constants which can be determined by evaluating  $\psi(m)$ at the limit masses.  We also set the metallicity of the Sun's birth cluster to $Z= 0.02$ $\left([Fe/H]=0 \right)$.

\begin{table}
 \centering
 \begin{minipage}{80mm}
  \caption{Radius (\rc), mass (\mc), number of particles ($N$) and velocity dispersion ($\sigma_\mathrm{v}$) adopted for the parental cluster of the Sun}
  \label{tab:ic_clusters}
  \begin{tabular}{@{} c c c c} \hline
  \textbf{\rc} \textbf{(pc)} & \textbf{\mc} $\mathbf{(M_{\odot}) }$ & $N$ & $\mathbf{\sigma_v (kms^{-1})}$ \\ \hline
   	 0.5& 510 & 875 & 2.91\\
	    1  & 641 &1050 & 2.29 \\
	       & 765 & 1050 & 2.27 \\
               & 1007 & 1741 & 2.96 \\
        1.5  & 525 & 875 & 1.61 \\
              &  1067 & 1740 & 2.42 \\
          2   & 1023 & 1741 & 2.12  \\
             & 883 & 1350 & 2.05  \\
         3   &  804 & 1500& 1.44 \\  \hline	
	\end{tabular}
\end{minipage}
\end{table}

\subsubsection{Primordial binary stars}

The dynamical evolution of stellar systems is affected by a non-negligible fraction of primordial
binaries (see e.g. \cite{tanikawa}). Therefore, we also modelled the Sun's birth cluster with
different primordial binary fractions in order to observe their effect on the current
phase-space distribution of the solar siblings. We varied the \binaryf\ from zero (only single
stars) up to 0.4.

We find that binaries have an effect on the internal evolution of the Sun's birth cluster, in the
sense that they tend to halt core collapse. The influence of binaries on the dissolution of siblings
throughout the Galactic disk is negligible. We observe that the current spatial distribution of the
solar siblings and their astrometric properties are little affected by the \binaryf\ of the Sun's
birth cluster. Thus hereafter we focus only on clusters with a \binaryf\ of zero.

\subsubsection{Initial phase-space coordinates }
\label{sect:initialps}

The initial centre of mass coordinates of the Sun's birth cluster \centermass\ were computed by
integrating the orbit of the Sun backwards in time taking into account the uncertainty in its
current Galactocentric position and velocity, using the same methods as \cite{martinezb14}. In these
simulations we ignore the vertical motion of the Sun.

We generate $5000$ random positions and velocities from a normal distribution centred at the current
Galactocentric phase-space coordinates of the Sun $(r_\odot, v_\odot)$.  Thus, the standard
deviations $(\sigma)$ of the normal distribution correspond to the measured uncertainties in these
coordinates.  We assume that the Sun is currently located at: $r_\odot= (-8.5,0, 0)$~kpc, with
$\sigma_r=(0.5, 0, 0)$~kpc. In this manner, the uncertainty in $y_\odot$ is set to zero given that
the Sun is located on the $x$-axis of the Galactic reference frame \citep[see e.g.][figure
1]{martinezb14}.

The present-day velocity of the Sun is $v_\odot= (U_\odot, V_\odot)$; where
\begin{align}
  U_\odot \pm \sigma_{U}  &= 11.1 \pm 1.2 \text{\ } \kms \nonumber \\
  V_\odot \pm \sigma_{V} &= (12.4+ V_\mathrm{LSR}) \pm 2.1 \text{\ }\kms\,.
\end{align}
Here, the vector $(11.1\pm1.2, 12.4\pm 2.1)$ \kms\ is the peculiar motion of the Sun
\citep{schonrich} and $V_\mathrm{LSR}$ is the velocity of the local standard of rest which depends
on the choice of Galactic parameters.

We integrate the orbit of the Sun backwards in time during $4.6$~Gyr, for each of the initial
conditions in the ensemble. At the end of the integration, we obtain a distribution of possible
phase-space coordinates of the Sun at birth  $\left(\pbirth \right)$.  This procedure was carried
out for 125 different Galactic parameters and models, according to the parameter value ranges listed
in Tables \ref{tab:galparams_p2} and \ref{tab:composite_p2}. We used 111 different combinations of
bar and spiral arm parameters for the 2 and 4-armed spiral models, and 14 different parameters for
the composite model.

Once the distribution \pbirth\ is obtained for a given galactic model we use the median of the
values of \pbirth\ as the value for \centermass. For the combinations of Galactic parameters used,
we found that the median value of \pbirth\ remains in the range of $8.5$--$9$~kpc. This is
consistent with \cite{martinezb14}, who found that the Sun hardly migrates in a Galactic potential
as the one explained in Sect.\ \ref{sect:Gmodel}. We therefore chose to fix
$||\vect{x}_\mathrm{cm}||= ||\vect{x}_\mathrm{b}||$ to a value of $9$~kpc, with the velocity
$\vect{v}_\mathrm{cm}$ corresponding to this value. We note that restricting the birth radius of the
Sun for a given Galactic model (fixed bar and spiral arm parameters) limits the possible outcomes
for the phase space distribution of the solar siblings. Different starting radii would lead to
different orbits which are affected differently by the bar and spiral arm potentials, which in turn
implies different predicted distributions of the solar siblings after $4.6$~Gyr.  Although we do not account for these differences in outcomes in our
simulations there is still significant spread in the predicted solar
sibling distribution caused by the different bar and
spiral arm parameters combinations we used (as demonstrated in Sect. \ref{sect:distrib}).

\subsection{Numerical simulations}
\label{sect:simulations}

The various simulation elements described above were to carry out simulations of the evolution of
the Sun's birth cluster as it orbits in the Milky Way potential. We used $9\times125= 1125$
different combinations of birth cluster and Galactic potential parameters, using the parameter
choices listed in tables \ref{tab:galparams_p2}, \ref{tab:composite_p2} and \ref{tab:ic_clusters},
in order to study a large variety of possible present-day phase space distributions of the solar
siblings.

We use the \textsc{huayno} code \citep{huayno} to compute the gravity among the stars within the
cluster. We set the time-step parameter to $\eta= 0.03$. We also use a softening length given by
\citep{aarseth}:
\begin{equation}
  \epsilon= \frac{4R_\mathrm{vir}}{N}\,,
\end{equation}
where $R_\mathrm{vir}$ is the initial virial radius of the cluster and $N$ the number of stars.

To calculate the external force due to the Galaxy we use a $6th$-order Rotating \textsc{bridge}
\citep[Pelupessy et al.\ in preparation;][]{martinezb14}. We set the \bridge\ time-step to
$dt= 0.5$~Myr\footnote{This set-up in the dynamical codes give a maximum energy error per time-step
in the simulations of the order of $10^{-7}$.}.

The stellar evolution effects were modelled with the population synthesis code \textsc{SeBa}
\citep{seba, seba1}. The magnitudes and colours of the stars were subsequently calculated from
synthetic spectral energy distributions corresponding to the present-day effective temperature and
surface gravity of the solar siblings. In addition the effects of extinction are accounted for. The
simulation of photometry is described further in Sect.\ \ref{sect:distrib}.

The various codes used to include the simulation elements above are all coupled through the {\amuse}
framework \citep{portegies13}.  In the simulations we evolve the Sun's birth cluster during
$4.6$~Gyr.

\begin{figure}[t]
  \centering
  \includegraphics[width= 7cm, height= 10.5cm]{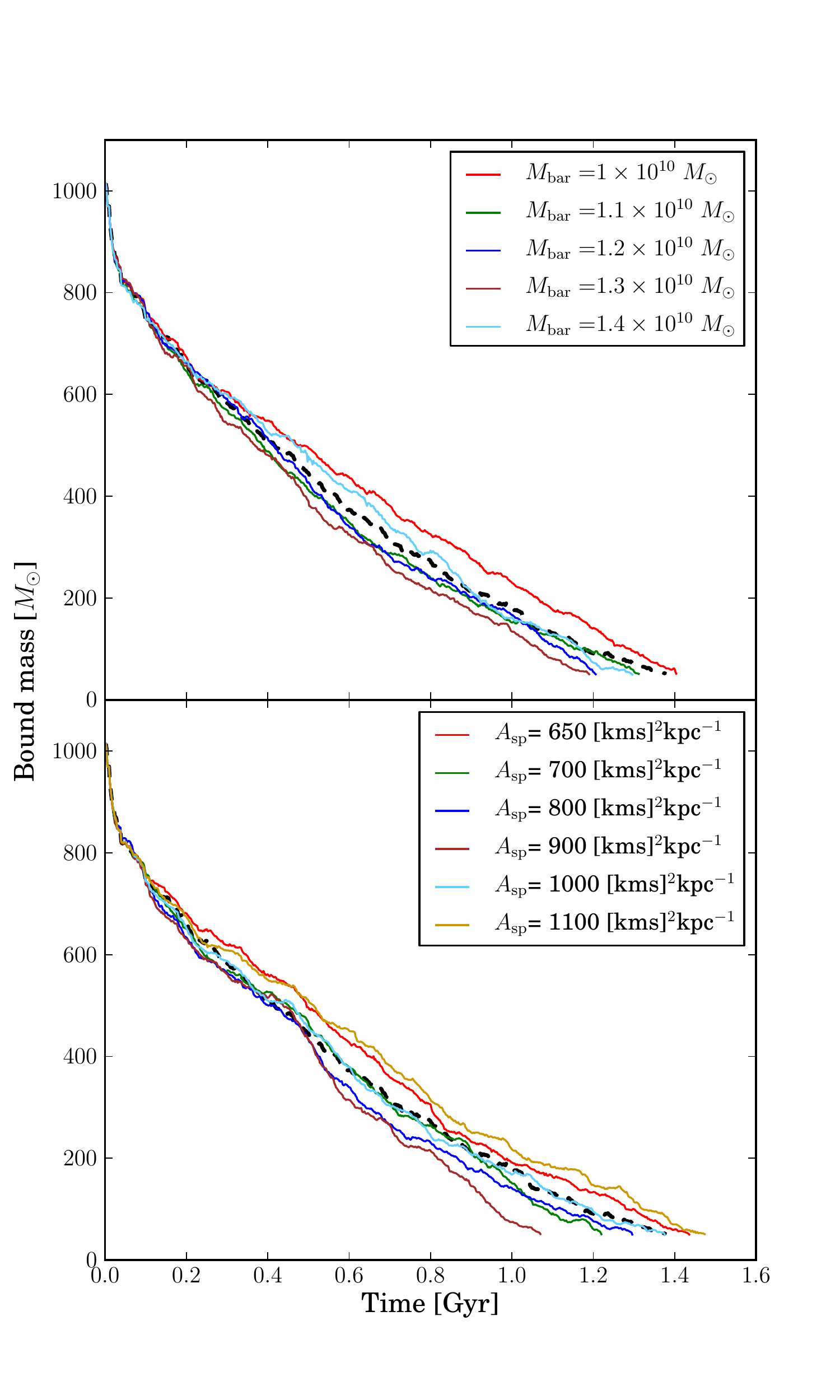}
  \caption{\textit{Top:} Bound mass of the Sun's birth cluster as a function of time for different
  masses of the central bar of the Galaxy. The dashed black line corresponds to the bound mass of
  the Sun's birth cluster for a purely axisymmetric Galactic model. \textit{Bottom:} Bound mass of
  the Sun's birth cluster as a function of time for different amplitudes of the spiral arms. The
  dashed black line has same meaning as above. Here the initial mass and radius of the
  Sun's birth cluster are $1023$~$M_\odot$ and $2$~pc respectively.
 \label{fig:diff_M}}
\end{figure}

\section{Disruption of the Sun's birth cluster}
\label{sect:ev_sbc}

As the Sun's birth cluster orbits in the Milky Way potential the tidal field and the effects of the
bar and spiral arms will cause the gradual dissolution of the cluster, its stars spreading out over
the Galactic disk. Here we briefly summarize our findings on the cluster dissolution times in our
simulations. The results are in line with what is already known about the dynamical evolution of
open clusters.

To compute the disruption rate of the Sun's birth cluster it is necessary to know its tidal radius
as a function of time. In its general form, the tidal radius is defined by the following expression
\citep{renaud, rieder}:
\begin{equation}
  \label{eq:rt}
  r_\mathrm{t}= \left( \frac{GM_\mathrm{c}}{\lambda_\mathrm{max}} \right)^{1/3}\,.
\end{equation}
Here $G$ is the gravitational constant, $M_\mathrm{c}$ is the mass of the cluster and
$\lambda_\mathrm{max}$ is the largest eigenvalue of the tidal tensor $T_{ij}$ which is defined as:
$T_{ij}=-\frac{\partial ^2\phi}{\partial x_i\partial x_j}$, with $\phi$ being the Galactic
potential.

We use the method of \cite{BM03} to compute the bound mass of the Sun's birth cluster iteratively.
At each time-step, we first assume that all stars are bound and we calculate the tidal radius of
the system through Eq.\ \ref{eq:rt}, using the value of $T_{ij}$ at the cluster centre. We use the
method of \cite{hop} to calculate the cluster centre. With this first estimate of $r_\mathrm{t}$ we
compute the bound mass, which is the mass of the stars that have a distance from the cluster centre
smaller than $r_\mathrm{t}$. We use this bound mass and the density centre of the bound particles to
recalculate $r_\mathrm{t}$ and make a final estimate of the bound mass. We consider the Sun's birth
cluster disrupted when $95\%$ of its initial mass is unbound from the cluster.

We studied the effect of the mass of the bar and the spiral arms on the cluster evolution by varying
the bar mass or the spiral arm strength, while keeping the other Galactic model parameters fixed.
The mass of the bar was varied for a fixed pattern speed of $\Omega_\mathrm{bar}=70$~\patspeed, and
with a fixed two-arm spiral with pattern speed $\Omega_\mathrm{sp}= 20$~\patspeed\ and amplitude
$A_\mathrm{sp}= 650$~\spiralampl. The effect of the spiral arm amplitude was studied for a two-arm
spiral with pattern speed $\Omega_\mathrm{sp}=18$~\patspeed, and a fixed bar with $M_\mathrm{bar}=
9.8\times10^9$~$M_\odot$ and $\Omega_\mathrm{bar}=40$~\patspeed. The resulting evolution of the
bound mass of the clusters is shown in Fig.\ \ref{fig:diff_M}, where the top panel shows the effect
of varying the bar mass and the bottom panel shows the effect of varying the spiral arm strength. In
both cases we also show the evolution for the case of a purely axisymmetric model of the Galaxy.

From Fig.\ \ref{fig:diff_M} is is clear that the disruption time of the cluster is not very
sensitive to the parameters of the Galactic model. The range of disruption times across all
our simulations is $0.5$--$2.3$~Gyr, with additional scatter introduced due to the different
perigalactica and eccentricities of the cluster orbits.

\section{Current distribution of Solar siblings in the Milky Way}
\label{sect:distrib}

\begin{figure}[t]
  \centering
  \includegraphics[width= 9cm, height= 9cm]{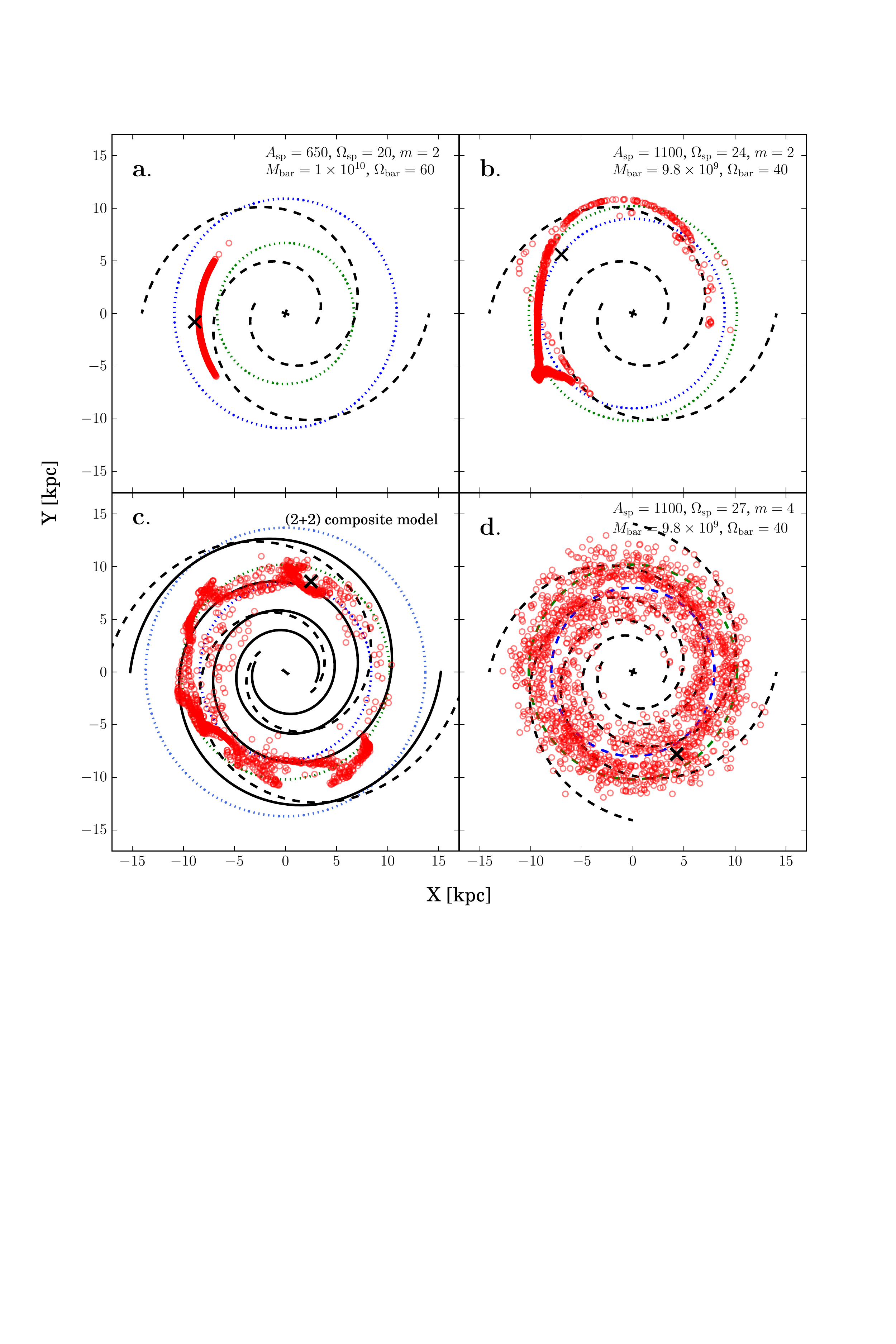}
  \caption{Present-day distribution of solar siblings in the $xy$ plane. The point $(0,0)$
  is the centre of the Milky Way. The dashed black lines represent the potential of the spiral arms
  at present. The dotted blue and green circles correspond to the \crsp\ and \olrbar\
  respectively. The black crosses in each panel mark the initial location of the Sun's birth
  cluster, which is at $9$~kpc. Here, the initial mass and radius of the Sun's birth cluster are
  $1023$~$M_\odot$ and $2$ parsec respectively. \textit{Top panels:} Distribution of solar siblings in a
  Galactic model with two spiral arms. The position of the \crsp\ and \olrbar\ are respectively:
  $(11, 6.7)$~kpc (\textbf{a}) and $(9, 10.2)$~kpc (\textbf{b}). \textit{Bottom panels:} \textbf{c.}
  Distribution of solar siblings in a (2+2) composite model with $A_\mathrm{sp1}=1300$~\spiralampl\
  . The solid and dashed black lines represent the main and secondary spiral structures with
  co-rotation resonances located at $8.4$ and $13.7$~kpc respectively. The \olrbar is at $10.2$~kpc.
  \textbf{d.} Distribution of solar siblings in a Galactic model with four spiral arms. The \crsp\
  and \olrbar\ are located at $8$ and $10.2$~kpc respectively. \label{fig:distrib_sib}}
\end{figure}

If the Sun's birth cluster was completely disrupted in the Galaxy at around $1.8$~Gyr, the Sun and
its siblings are currently spread out over the Galactic disk, since they have been going around the
Galaxy on individual orbits during the last $2.8$~Gyr. In Fig.\ \ref{fig:distrib_sib} we show four
possible distributions of the solar siblings in the Galactic disk. Note that in contrast to the
cluster disruption time, the present-day distribution of solar siblings depends strongly on the
Galactic parameters, especially on changes in $m$, $\Omega_\mathrm{sp}$ and $\Omega_\mathrm{bar}$.
This is because the motion of the solar siblings depends on whether their orbits are affected by the
\crsp\ or by the \olrbar. For instance, in panel \textbf{a} of Fig.\ \ref{fig:distrib_sib} we
observe that there is not much radial migration with respect to the initial position of the Sun's
birth cluster ($\bar{R}_\mathrm{sib}-R_\mathrm{i} \sim 0.5$~kpc, where
$R_\mathrm{i}=||\vect{x}_\mathrm{cm}||$). In this example, the Sun and its siblings are not
considerably influenced by the \crsp\ or by the \olrbar\ during their motion in the Galactic disk.
The apocentre and pericentre of the solar siblings is at around $7$ and $10$~kpc; while the \crsp\
and \olrbar\ are located at $11$ and $6.7$~kpc respectively. This distribution of solar siblings is
similar to the distributions predicted by \cite{portegies09} and \cite{brown10}.

\begin{figure}
  \centering
  \includegraphics[width= 9cm, height= 12cm]{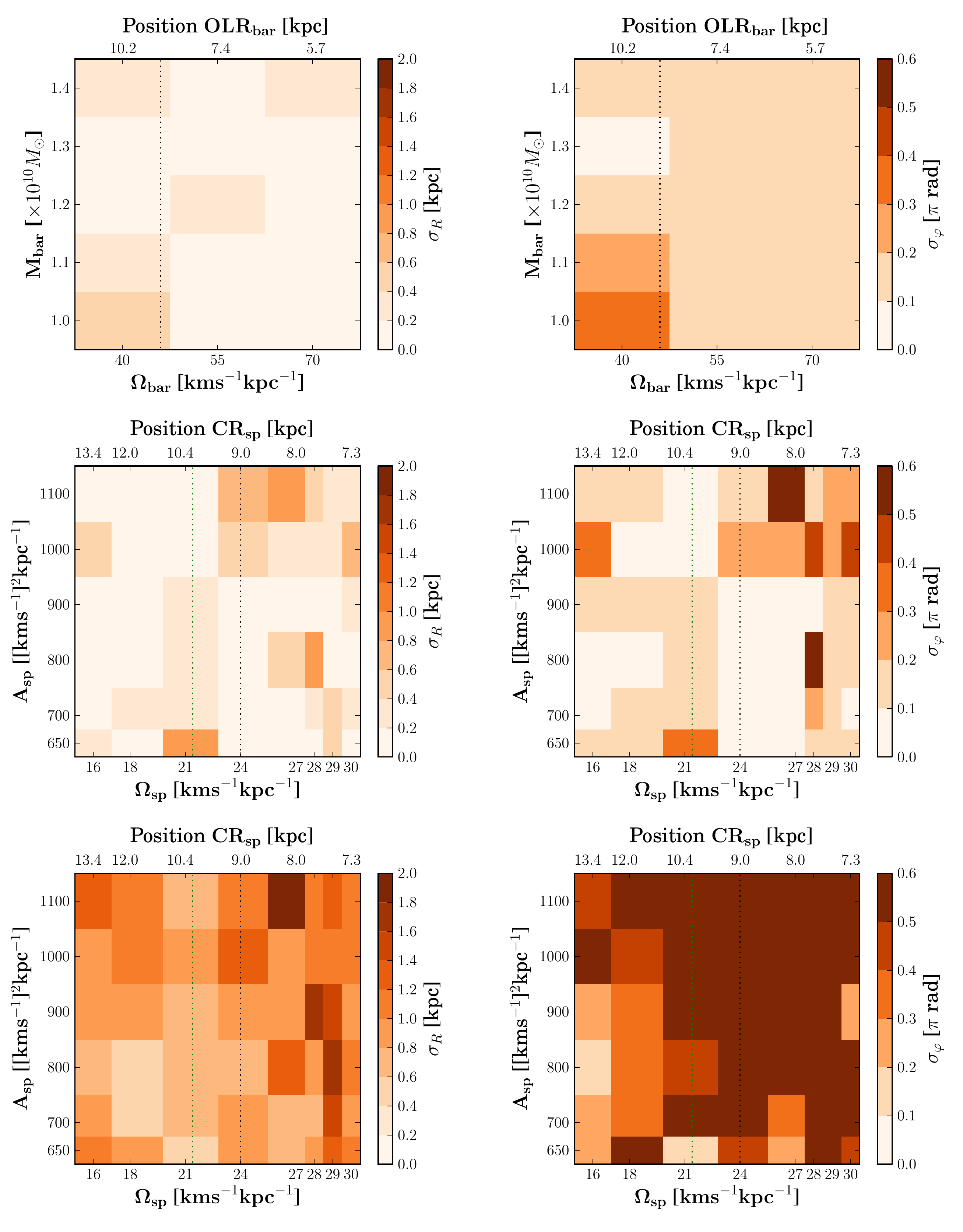}
  \caption{Radial and angular dispersion of the current distribution of solar siblings as a function
  of different Galactic parameters. \textit{Top:} The mass and pattern speed of the bar are varied.
  Here $A_\mathrm{sp}=650$~\spiralampl, $\Omega_\mathrm{sp}=20$~\patspeed\ and $m=2$.
  \textit{Middle:} The amplitude and pattern speed of the spiral structure changes. The Galaxy has
  two spiral arms. \textit{Bottom:} The same as in the middle panel but for a Galaxy with four
  spiral arms. In the Middle and bottom panels, $M_\mathrm{bar}= 9.8\times10^9$~$M_\odot$ and
  $\Omega_\mathrm{bar}=40$~\patspeed. For this set of simulations $\mc=1023$~$M_\odot$ and
  $\rc=2$~pc. The dotted black line in the panels corresponds to $||\vect{x}_\mathrm{cm}||$. The
  dotted green line in the middle and bottom panels represents the \olrbar\ which is located at
  $10.2$~kpc from the Galactic centre. In the top panel the value of \crsp\ is fixed at $10.9$~kpc.
  \label{fig:distrib_GP}}
\end{figure}

If the \crsp\ and the \olrbar\ are located in the same region where the Sun and its siblings move
around the Galaxy, these stars will undergo constant and sudden changes in their angular momentum.
As a consequence, the distribution of solar siblings will contain lots of substructures. This effect
can be observed in panels \textbf{b} and \textbf{c} of Fig.\ \ref{fig:distrib_sib}.

When the Sun's birth cluster evolves in a Galaxy containing four spiral arms, the solar siblings
undergo considerable radial migration. As a consequence, the current distribution of solar siblings
is highly dispersed in galactocentric radius and azimuth, as observed in panel \textbf{d} of Fig.\
\ref{fig:distrib_sib}. In this Galactic environment, some solar siblings can be located at radial
distances of up to $3$~kpc different from the radial distance of the Sun to the Galactic centre.

\cite{mishurov11} presented the spatial distribution of solar siblings in a Galactic potential with
transient spiral structure of different life-times. They found that the solar siblings are dispersed
all over the disk. Some of these stars can be even located at distances larger than $10$~kpc with
respect to the Galactic centre (see Figs. $9$ and $10$ in their paper). By comparing these results
with the distributions that we obtained for a four-armed spiral structure (panel \textbf{d} Fig.\
\ref{fig:distrib_sib}), we infer that the solar siblings would be even more dispersed and located
farther from the Sun if the spiral structure of the Milky Way were transient.

\cite{BH10} used stellar diffusion modelling to predict the current distribution of solar siblings
in the Galaxy. They used four different approaches, starting from constant and isotropic
coefficients to models where they accounted for the impact of churning on the solar siblings. In
their approach the solar siblings are always spread all over the Galactic disk (all azimuths), in a
configuration like the one shown in Fig.\ \ref{fig:distrib_sib}\textbf{d}. None of their solar
siblings distributions show substructures or stellar concentrations in radius and azimuth, as is
shown in Figs. \ref{fig:distrib_sib}\textbf{a}--\textbf{c}. \cite{BH10} found that a substantial
fraction of solar siblings may be located at galactic longitudes of $l=90^\circ$--$120^\circ$ or
$l=30^\circ$--$60^\circ$, depending on the diffusion model employed.

We characterize our predicted present-day distributions of solar siblings by means of their radial
and azimuthal dispersion ($\sigma_\mathrm{R}$ and $\sigma_\phi$). These quantities are computed
using the Robust Scatter Estimate (RSE) \citep{lindegren}. The radial dispersion of the
distributions shown in panels \textbf{a}--\textbf{d} in Fig.\ \ref{fig:distrib_sib} are
$\sigma_\mathrm{R}= 0.1$, $0.4$, $0.9$, and $1.8$~kpc, respectively. The angular dispersion of these
distributions is: $\sigma_\phi= 0.1\pi$, $0.2\pi$, $0.4\pi$, and $0.6\pi$ rad. Since $0.6\pi$
corresponds to the standard deviation of a uniform distribution in azimuth, a highly dispersed
distribution (as in panel \textbf{d} of Fig.\ \ref{fig:distrib_sib}) satisfies $\sigma_\mathrm{R} >
0.9$~kpc and $\sigma_\phi > 0.4\pi$~rad.

In Fig.\ \ref{fig:distrib_GP} we show the radial and angular dispersion of the current distribution
of solar siblings as a function of different Galactic parameters. In the top panel we varied the
parameters of the bar. In the middle and bottom panels, we varied the amplitude and pattern speed of
the spiral arms. Note that there is a remarkable increase in $\sigma_\mathrm{R}$ and $\sigma_\phi$
when the Galaxy has four spiral arms. In that Galactic potential, $83$\% of the simulations result
in the solar siblings currently being dispersed all over the Galactic disk ($\sigma_\mathrm{R} >
0.9$~kpc and $\sigma_\phi > 0.4\pi$~rad). On the contrary, in a Galaxy with two spiral arms (e.g.\
Fig.\ \ref{fig:distrib_GP}, top and middle panels), the spatial distribution of solar siblings is
more `clustered' in radius and azimuth. We found that in $84$\% of these simulations,
$\sigma_\mathrm{R} < 0.4$~kpc and $\sigma_\phi < 0.2\pi$~rad.

We computed $\sigma_\mathrm{R}$ and $\sigma_\phi$ for different initial conditions of the Sun's
birth cluster, according to the values presented in table \ref{tab:ic_clusters}. We found that
$\sigma_\mathrm{R}$ and $\sigma_\phi$ do not depend on $M_\mathrm{c}$ and $R_\mathrm{c}$. The
maximum difference in radial and angular dispersion is $\Delta \sigma_\mathrm{R_{max}}= 0.2$~kpc and
$\Delta\sigma_{\phi_\mathrm{max}}= 0.2\pi$ rad.

The current distribution of solar siblings constrains the number of stars that can be observed near
the Sun. For instance, if the solar siblings are `clustered' in galactocentric radius and azimuth
(as shown at the top and middle panels of Fig.\ \ref{fig:distrib_GP}), the probability of finding a
large fraction of solar siblings in the vicinity of the Sun increases. Conversely, in more dispersed
solar siblings distributions (e.g.\ bottom panel Fig.\ \ref{fig:distrib_GP}), we expect to find a
smaller fraction of solar siblings in the solar vicinity.

We next consider the prospects of identifying solar sibling candidates from the future \gaia\
catalogue data.

\section{The search for the solar siblings with \gaia}
\label{sect:kss}

The \gaia\ mission will provide an astrometric and photometric survey of more than one billion stars
brighter than magnitude $G=20$ \citep{gaia}, where $G$ denotes the apparent magnitude in the white
light band of used for the astrometric measurements, covering the wavelength range $\sim350$--$1050$
nm \citep[see][]{jordi10}. Parallaxes ($\varpi$) and proper motions ($\mu$) will be measured with
accuracies ranging from $10$ to $30$ micro-arcseconds ($\mu$as) for stars brighter than $15$ mag,
and from $130$ to $600$ $\mu$as for sources at $G=20$. For $\sim100$ million stars brighter than
$G=16$ \gaia\ will also measure radial velocities ($V_r$), with accuracies ranging from $1$ to $15$
\velocity. \gaia\ will not only revolutionize the current view of the Galaxy but will generate a
data set which should in principle allow for a search for solar siblings even far away from the Sun.

In this section we use our simulations to predict the number of solar siblings that will be seen by
\gaia, and to study their distribution in the space of parallax, proper motion, and radial velocity
with the aim of establishing efficient ways of selecting solar sibling candidates from the \gaia\
catalogue.

\subsection{The solar siblings in the \gaia\ catalogue}

We first compare the predicted \gaia\ survey of the solar siblings with predictions by \cite{BH10},
who considered the prospects for a survey like \galah\ \citep{galah} to varying limiting magnitudes.
Following \cite{BH10} we broadly distinguish the possible present-day phase configurations for the
solar siblings by referring to the cases shown in the panels of Fig.\ \ref{fig:distrib_sib} as
\modela\ and \modelb\ (compact spatial distribution of solar siblings), \modelc\ (spatial
distribution of solar siblings obtained with the $2+2$ composite model) and \modeld\ (highly
dispersed spatial distribution of solar siblings).

In predicting the observed kinematic properties of the solar siblings we want to account for the
fact that we do not know which of the stars in our simulated clusters is the Sun. The location of
the Sun with respect to its siblings will affect the number of siblings that can be observed,
especially for clusters that during their dissolution have not spread all over the Galactic disk in
azimuth. We therefore proceed as follows. All stars in the simulated cluster located at
Galactocentric distances of $R=8$--$9$ kpc and with stellar masses around $1$~$M_\odot$ are
considered possible `suns'. The \gaia\ observables $(\varpi, \mu, V_\mathrm{r})$ of the siblings are
then calculated with respect to each of these candidate suns. This results in a set of distributions
of siblings over the observables which can be considered collectively in order to account for the
uncertain position of the Sun within its dissolved birth cluster.

We used the \textsc{PyGaia}\footnote{\textit{https://pypi.python.org/pypi/PyGaia/}} code to compute
the astrometric properties of the solar siblings. Since we are interested in solar siblings that can
be observed by \gaia, we only include stars for which $G\leq 20$.

\begin{figure}
  \centering
  \includegraphics[width= 8.5 cm, height= 9cm]{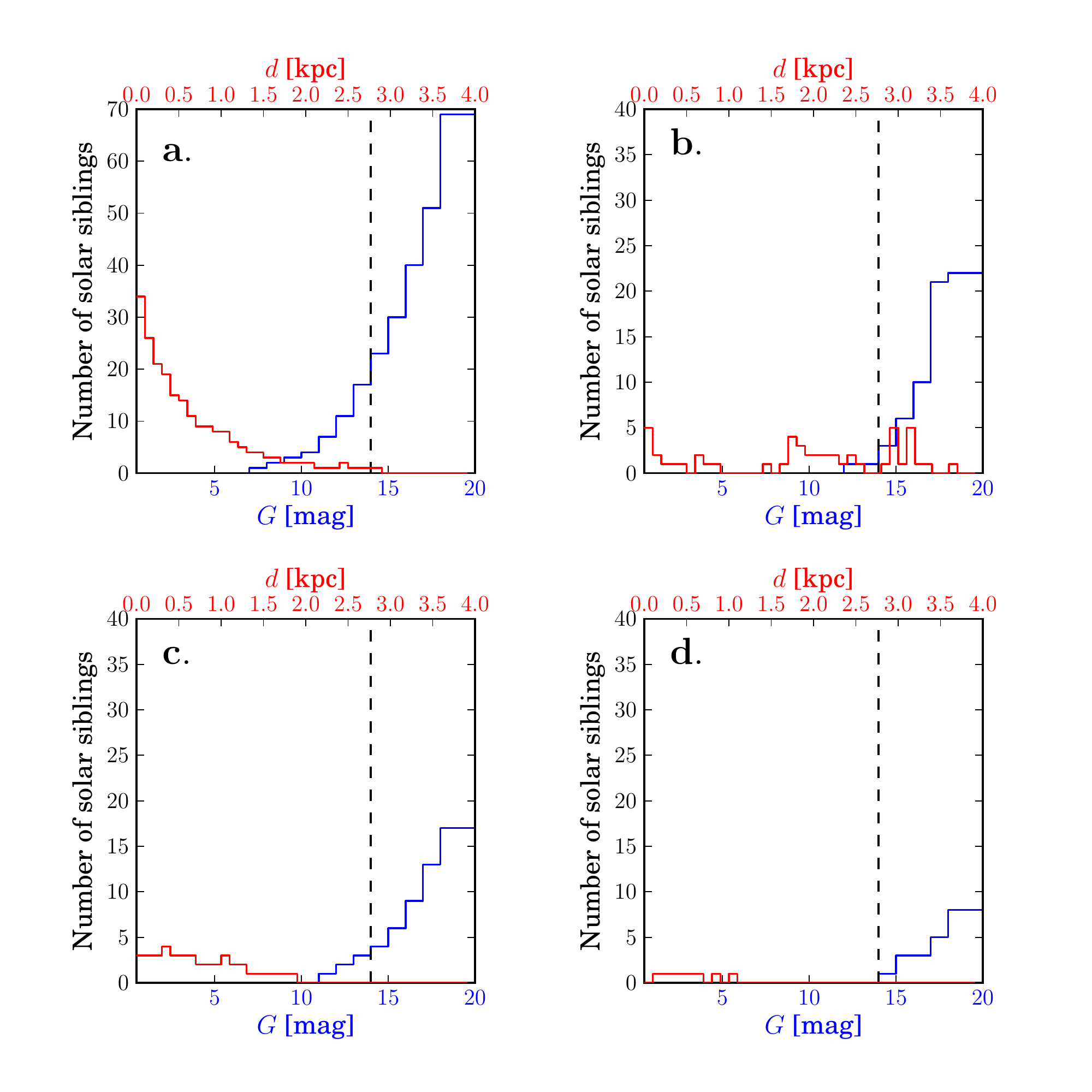}
  \caption{Median number of solar siblings that \gaia\ is predicted to observe, as a function of
    their heliocentric distances $d$ (red histograms) and $G$ magnitudes (blue histograms). The
    letters in the left corner correspond to the distributions shown in Fig.\ \ref{fig:distrib_sib}.
    The vertical dotted black lines in each panel represent the limiting magnitude of the \galah\
    survey, $G\sim14$ mag. \label{fig:num_ss}}
\end{figure}

The apparent $G$ magnitude is given by the following equation \citep{jordi10}:
\begin{equation}
  G= -2.5\log{\left( \frac{\bigintss_{\lambda_\mathrm{min}}^{\lambda_\mathrm{max}}
  F(\lambda)10^{-0.4A_\lambda}S_x(\lambda)
  d\lambda}{\bigintss_{\lambda_\mathrm{min}}^{\lambda_\mathrm{max}} F^\mathrm{Vega}(\lambda)
  S_x(\lambda) d\lambda}  \right)} + G^\mathrm{Vega}\,. \label{eq:mag}
\end{equation}
Here $F(\lambda)$ and  $F^\mathrm{Vega}(\lambda)$ are the fluxes of a solar sibling and Vega,
respectively, as measured above the atmosphere of the Earth (in photons~s$^{-1}$~nm$^{-1}$).  We
obtain $F(\lambda)$ through the BaSeL library of synthetic spectra \citep{basel}, by searching for
the stellar spectral energy distribution which best matches the mass ($M_\mathrm{s}$), radius
($R_\mathrm{s}$)  and effective temperature  ($T_\mathrm{eff}$) of a given solar sibling, where the
latter quantities are obtained from the stellar evolution part of the simulations.
$F^\mathrm{Vega}(\lambda)$ was obtained in the same way by using the following parameters
\citep{jordi10}: $T_\mathrm{eff}=9550$~K, $\log g =3.95$~dex, $\mathrm{[Fe/H]}= -0.5$~dex and
$\epsilon_t= 2$~\kms.

$A_\lambda$ in Eq.\ \ref{eq:mag} is the extinction, which is described by:

\begin{equation}
  A_\lambda= A_\mathrm{V}\left( a_\lambda +\frac{b_\lambda}{R_\mathrm{V}} \right)\,,
\end{equation}
where $A_\mathrm{V}$ is the extinction in the visual (at $\lambda= 550$~nm).  The value of
$A_\mathrm{V}$ within our simulated Galaxy is computed by means of the Drimmel extinction model
\citep{drimmel_model}. $R_\mathrm{V}$ is the ratio between the extinction and colour excess in the
visual band; we use $R_\mathrm{V}=3.1$. $a_\lambda$ and $b_\lambda$ are coefficients calculated
trough the Cardelli extinction law \citep{cardelli}.

The function $S_x(\lambda)$ in Eq.\ \ref{eq:mag}  corresponds to the \gaia\ pass-bands, which depend
on the telescope transmission and the CCD quantum efficiency. To compute the stellar magnitude in
$G$, we use the corresponding pass-band described in \cite{jordi10}.

Finally,  $G^\mathrm{Vega}$ is the magnitude zero point which is fixed through the measurement of
the flux of Vega, such that  $G^\mathrm{Vega}= 0.03$~mag.

\begin{table}
 \centering
 \begin{minipage}{90mm}
  \caption{Median and RSE of the number of solar siblings observed at different
    heliocentric distances and to different limits in $G$. The last column lists the total number of
    solar siblings out to the magnitude limit listed. The first column refers to the distributions
    shown in Fig.\ \ref{fig:distrib_sib}. The statistics for a given model were
    obtained from the distribution of the number of observable solar siblings predicted for each of
  the candidate Suns. \label{tab:ss}}
  \begin{tabular}{@{} c c c c c c} \hline
Model & $G$ [mag] & $d\leq 100$~pc & $d\leq 500$~pc & $d\leq 1$~kpc & total \\ \hline
 \textbf{a} & $\leq 14$ & $14\pm 5$ & $26\pm7$ & $30\pm7$ & $31\pm7$ \\
	 & $\leq 16$        & $22 \pm 8$ & $50\pm16$ & $62\pm18$ & $72\pm19$ \\
	 & $\leq 18$        & $31\pm 13$ & $95\pm33$ & $121\pm39$ & $146\pm38$ \\
	 & $\leq 20$        & $33\pm 14$ &$145\pm49$ & $199\pm62$ & $268\pm57$ \\ \hline
\textbf{b} & $\leq 14$  & $1\pm0.3$ & $1\pm 0.6$ & $1\pm0.6$ & $1\pm0.6$ \\	
          & $\leq 16$       & $1\pm 0.9$ & $3\pm1$ & $3\pm1$ & $4\pm1$ \\
          & $\leq 18$       & $3\pm 2$ & $8\pm4$ & $10\pm6$ & $19\pm2$ \\
          & $\leq 20$       & $5 \pm 3$ & $14\pm8$ & $19\pm11$ & $61\pm0.3$ \\ \hline
\textbf{c} & $\leq 14$  & $1\pm1$ & $4\pm2$ & $5\pm3$ & $6\pm3$ \\	
	 & $\leq 16$       & $1\pm1$ & $8\pm4$ & $11\pm5$ & $15\pm6$ \\
	  & $\leq 18$      & $2\pm2$ & $13\pm7$ & $19\pm11$ & $33\pm16$ \\
	 & $\leq 20$       & $2\pm2$ & $18\pm10$ & $37\pm18$ & $61\pm31$ \\ \hline
\textbf{d} & $\leq 14$ & $0$ & $0$           & $1\pm0.7$ & $1\pm1$ \\	
	 & $\leq 16$      & $0$ & $1\pm1$ & $2\pm1$ & $4\pm1$ \\
	 & $\leq 18$      & $0$ & $2\pm1$ & $4\pm1$ & $9\pm2$ \\
	 & $\leq 20$      & $0$ & $4\pm1$ & $10\pm2$ & $22\pm4$ \\ \hline	 
\end{tabular}
\end{minipage}
\end{table}


In Fig.\ \ref{fig:num_ss} and Table \ref{tab:ss} we show the number of solar siblings that might be
observed by \gaia\ as a function of their heliocentric distances $d$ and their magnitudes $G$, where
we have averaged over each of the candidate Suns per model. Note that for models \textbf{a},
\textbf{c} and \textbf{d} the largest fraction of solar siblings is located within $\sim 500$~pc
from the Sun. Yet, the number of solar siblings located at this distance is rather small for some
cases. In models \textbf{c} and \textbf{d} for instance, just $18$ and $4$ solar siblings are at
$d\leq500$~pc on average (see table \ref{tab:ss}). In model \textbf{a}, on the other hand,
$145\pm49$ solar siblings might be identified. In model \textbf{b} the solar siblings are almost
uniformly distributed throughout the entire range of $d$, with more stars at $1.5 \lesssim d\lesssim
3.3$~kpc. A closer look at Fig.\ \ref{fig:num_ss} (and also at table \ref{tab:ss}) reveals that only
in the most 'clustered' spatial distribution of solar siblings (\modela) there is a chance to
observe tens of solar siblings within $100$~pc from the Sun, in accordance with \cite{portegies09}
and \cite{valtonen15}. In \modeld, on the contrary, it is not possible to observe substantial
numbers of solar siblings near the Sun.

Similar predictions of the observable number of solar siblings were made by \cite{BH10} in the
context of preparations for chemical tagging surveys, (their table 1). They assumed a larger birth
cluster of the Sun (with $2\times10^4$ stars) with a slightly more massive lower limit on the IMF
($0.15$~$M_\odot$ vs.\ $0.08$~$M_\odot$ in our case).

\subsection{Selecting solar sibling candidates from the \gaia\ catalogue}

\cite{brown10} used their simulated distribution of solar siblings to propose a criterion for the
selection of solar sibling candidates on the basis of their observed parallax and proper motion.
They basically proposed to select nearby stars with small motions with respect to the Sun. This
was motivated by the observation that in that region of the parallax vs.\ proper motion plane the ratio
between the number of siblings and the number of disk stars (in the Hipparcos catalogue) was
largest. Given that this contrast between the number solar siblings and disk stars depends on the
details of the Galactic potential (as illustrated in Fig.\ \ref{fig:distrib_sib}) we revisit the
selection criterion proposed by \cite{brown10} in order to assess how robust it is against the
uncertainties in the present-day distribution of solar siblings. We proceed in a similar way as
\cite{brown10} and examine the simulated present-day distribution of solar siblings in the space of
the astrometric observables (parallax, proper motion, radial velocity), and compare that to the
distribution of disk stars. We then search for regions in $(\varpi,\mu,V_\mathrm{r})$ where the
contrast between solar siblings and disk stars is high.

\begin{figure}
  \centering
  \includegraphics[width= 8.5cm, height= 9cm]{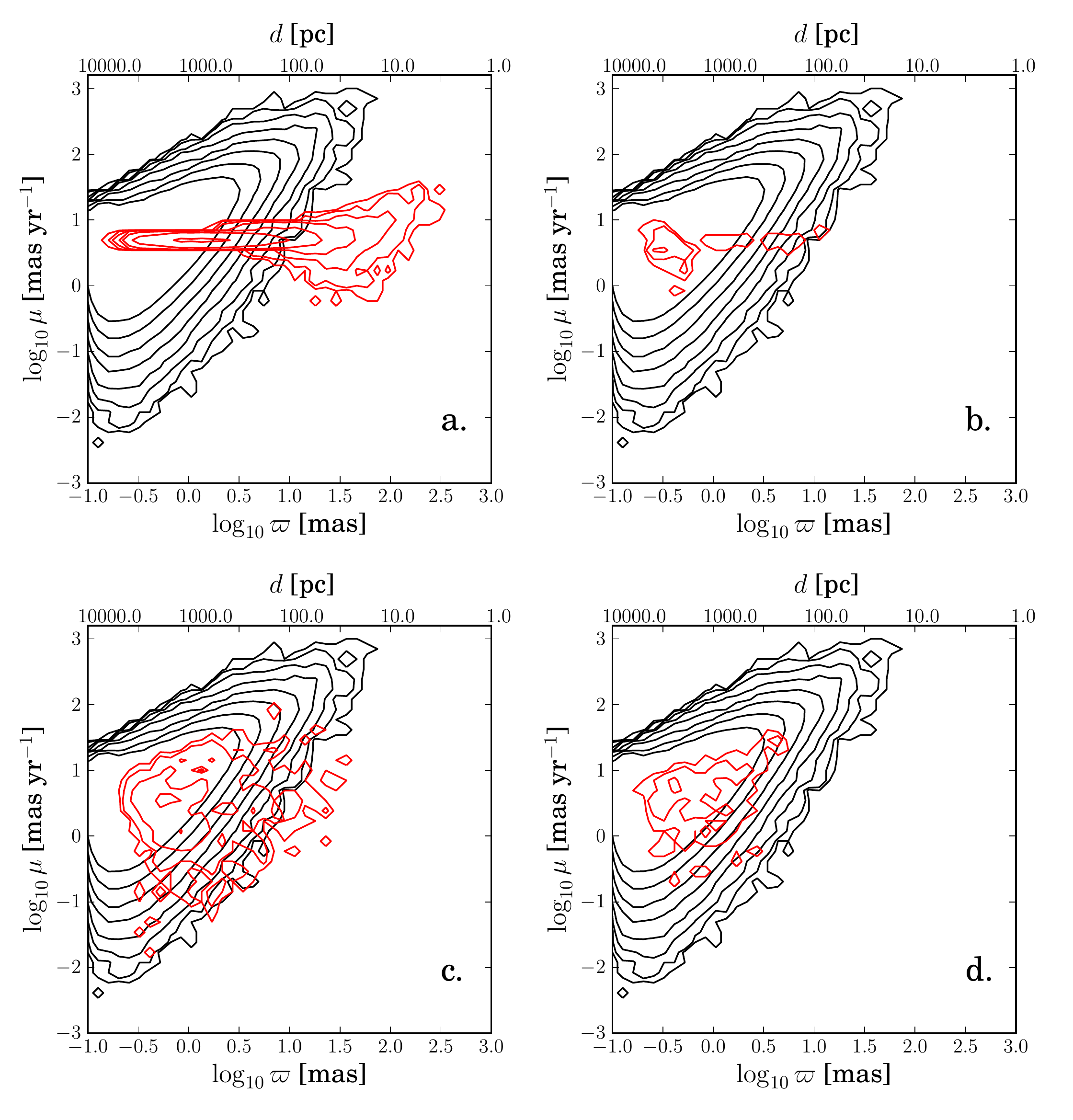}
  \caption{Distribution of solar siblings (red contours) and simulated \gaia\ data for disk stars
  (black contours) in the proper motion-parallax plane. Each panel corresponds to the distributions
  shown in Fig.\ \ref{fig:distrib_sib}. The red and black contours indicate the number of stars in
  bins of $0.1\times0.15$~mas$^2$yr$^{-1}$. The contour levels are at $1$, $3$, $10$, $30$, $100$,
  $300$, $1000$ and $3000$ stars/bin. In the labels of the top, we also show the heliocentric
  distance corresponding to each parallax. The proper motion axis represents to total proper motion
  of the star.\label{fig:plxpm}}
\end{figure}

We illustrate this procedure in Fig.\ \ref{fig:plxpm}. Here, the distribution of solar siblings in
the proper motion-parallax plane is represented by the red contours. The black contours correspond
to a simulation of field disk stars as measured by \gaia. We use the \gaia\ Universe Model Snapshot
(GUMS) \citep{gums} to generate a simulated sample of $2.6\times10^7$ field disk stars. GUMS
represents a synthetic catalogue of stars that simulates what \gaia\ will observe. To select only
disk stars, we used only the GUMS stars located in a cylindrical region of radius $8$~kpc and height
$300$~pc (i.e.\ $|z|\leq150$~pc) centred on to the Sun. The GUMS model includes multiple-star
systems. We determine which ones will be resolved by \gaia\ by using a prescription employed within
the Data Processing and Analysis Consortium
\citep[DPAC,][]{mignard}\footnote{http://www.cosmos.esa.int/web/gaia/dpac}. In this approach the
angular separation on the sky that \gaia\ can resolve depends on the apparent magnitudes of the
stars in the system, with the minimum separation being $\sim 38$~mas. For the unresolved cases, a
single detection is considered by computing the total integrated magnitude and averaging positions
and velocities.

\begin{figure*}
  \centering
  \includegraphics[width= 18cm, height= 15cm]{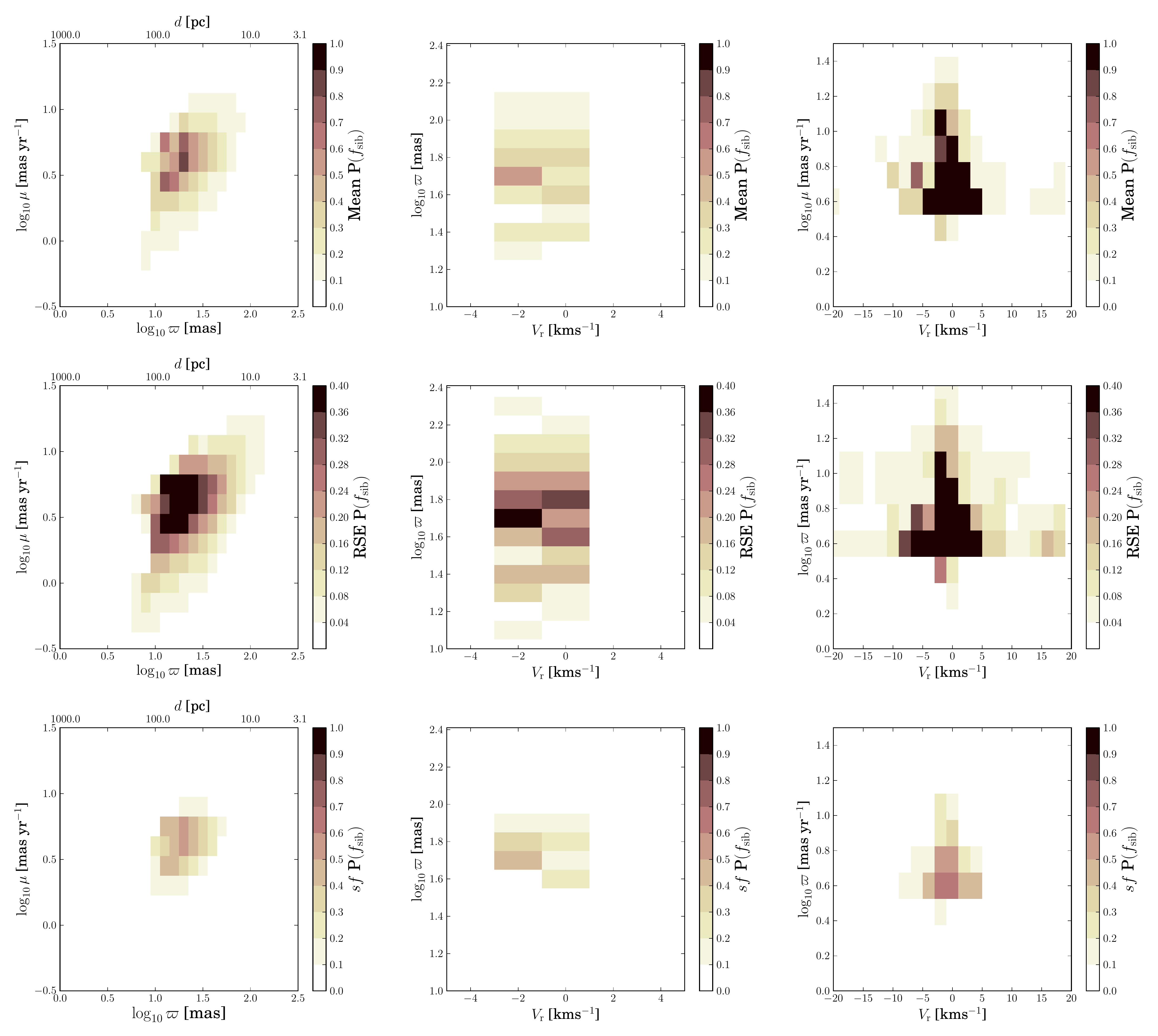}
  \caption{ Mean (top), RSE (middle) and survival function (bottom) of P$(f_\mathrm{sib})$ (see
  text). We show the projections of such a distribution in the proper motion versus parallax plane
  (left), in the parallax versus radial velocity plane (middle) and in the proper motion versus
  radial velocity plane (right). The bin area in each column is $(0.1\times0.15)$~mas$^2$~yr$^{-1}$,
  $(2\times0.15)$~\kms~mas and $(2\times0.1)$~\kms~\maspyr respectively.
  \label{fig:siblingness}}
\end{figure*}

As can be seen in Fig.\ \ref{fig:plxpm}, most of the solar siblings are located well within the
overall disk population (at distances over 100 pc) making the selection of sibling candidates on the
basis of astrometric and radial velocity data alone very difficult. The only area where a high
contrast between the number of siblings and disk stars can be expected is at large parallax and
small proper motion values. However, and as expected, this contrast depends strongly on the Galactic
potential used in predicting the solar sibling distribution. In order to evaluate the robustness of
a selection of sibling candidates in $(\varpi,\mu,V_\mathrm{r})$ we must take the uncertainties in
their distribution into account and we proceed as follows.

We divide the space $\varpi$, $\mu$ and $V_\mathrm{r}$ into discrete (3D) bins and determine for a
given simulated solar sibling distribution the number of solar siblings $N_\mathrm{sib}$ in each
bin. We also determine the number of disk stars $N_\mathrm{disk}$ in each bin and then calculate the
number $\SF= N_\mathrm{sib}/N_\mathrm{disk}$, which we refer to as the sibling fraction. The idea is
that a high value of \SF\ (say $\SF>0.5$) suggests that selecting stars from the corresponding
$(\varpi,\mu,V_\mathrm{r})$ bin in the Gaia catalogue should increase the success rate of subsequent
searches for solar siblings that examine the astrophysical properties of those stars (age,
metallicity, chemical abundance pattern). Alternatively the number \SF\ can be interpreted as
meaning that a star selected from the corresponding bin in $(\varpi,\mu,V_\mathrm{r})$ has a
probability \SF\ of being a solar sibling (provided of course that the simulated population of
siblings and disk stars is representative of reality).

To account for the uncertainties in the phase space distribution of siblings we repeat the above
procedure for each of our 1125 simulated solar sibling populations and for each of the `suns' within
a given population of siblings. This leads to a distribution of values of \SF, $p(\SF)$, for each
bin in $(\varpi,\mu,V_\mathrm{r})$. This distribution thus reflects different Galactic potential
parameters, different initial conditions for the Sun's birth cluster, and different possible
locations of the Sun within the dispersed sibling population. In Fig.\ \ref{fig:siblingness} we show
the mean value (top panel), the RSE (middle panel) and the survival function ($S(0.5)$) (bottom
panel) of $p(\SF)$. The survival function corresponds to the fraction of simulations for which
$\SF>0.5$, which provides a more robust indication of bins in $(\varpi,\mu,V_\mathrm{r})$ where a
high fraction of solar siblings is likely to be found. Note that the figure shows the statistics for
$p(\SF)$ marginalized over the coordinate not included in the plot.

The statistics of \SF\ shown in  Fig.\ \ref{fig:siblingness} show that the proposal by
\cite{brown10}, to search for solar siblings among nearby stars with small motions with respect to
the Sun, is robust to the uncertainties in the distribution of the solar siblings due to the
uncertain Galactic potential and birth cluster conditions. By examining the
$(\varpi,\mu,V_\mathrm{r})$ in three dimensions and looking for regions where the mean of $p(\SF)$
is above $0.5$, we refine the solar sibling candidate selection criterion by \cite{brown10} to:
\begin{align}
  \varpi & \geq 5~\text{mas}; \nonumber \\
  4 \leq \mu  &\leq 6~\text{mas yr}^{-1}; \nonumber \\
  -2 \leq V_\mathrm{r} & \leq 0~\kms. \label{eq:SScriterion}
\end{align}
The survival function in this region goes from $0.42$ to $0.54$. This indicates that despite the
uncertainties in the spatial distributions of solar siblings it is still possible to identify
regions in the space of $\varpi,\mu$ and $V_\mathrm{r}$ where more than a half of the stars might be
a solar sibling.

\section{Discussion}
\label{sect:discuss}

\subsection{Re-evaluation of existing solar sibling candidates}

\begin{table*}
 \centering
 \begin{minipage}{200mm}
  \caption{Current Solar siblings candidates. They are sorted by the value of $\SF$.}
  \label{tab:SS_candidates}
  \begin{tabular}{l c c c c c c c c c c c c}
 \\ \hline\noalign{\smallskip}
  Star name & d & $\sigma_\mathrm{d}$ & $\varpi$ & $\sigma_\varpi$& $\mu$ & $\sigma_\mu$ & $V\mathrm{r}$ & $\sigma_{V_\mathrm{r}}$& \SF\ &  RSE & $sf$& Ref.$^1$ \\
    (HD no.)    & (pc)& (pc)& (mas) &  (mas) & (\maspyr) &(\maspyr) &(\kms) &  (\kms) & & &  \\\hline\noalign{\smallskip}
  $147443$&$92.0$&$8.38$&$10.87$& $0.99$&$5.26$&$0.69 $&$-2.1$&$7.1$&$0.76$& $0.20$& $0.47$ &Br10 \\
  $196676$ &$74.4$  &$2.77$ &$13.44$& $0.5$& $5.06$&$0.54$&$-0.79$ &$0.1$& $0.56$& $0.38$& 0.42 &Br10 \\
  $192324$ &$67.11$& $4.82$ &$14.9$&  $1.07$&$6.36$& $2.01$& $-4.4$&$0.4$& $0.02$& $0.01$& 0.01 &Br10\\
  $46301$ &     $107.64$& $ 6.6$ &      $9.29$& $0.57 $&   $5.85$&   $0.71$&      $-6.7$ &$ 0.7 $&$0.01$& $0.005$& $0.01$&Ba12\\
  $162826 $ &   $33.6$ &  $0.41$ &        $29.76$&  $0.36 $&   $20.14$& $0.38 $&    $1.88$ &$ 0.0063$& $0.003$& $0.001$&$\sim 10^{-4}$ &Bo11\\
  $26690$ & $36.34$& $0.77$ &$27.52$&$0.58$&$3.62$&$0.58$&$2.4$&$1.9$&$0.003$&$0.001$& $\sim 10^{-4}$& Ba12 \\
  $207164$ &   $76.1$ &  $3.82  $&       $13.14$&   $0.66 $&   $3.06$& $0.7 $&        $-7.0 $ &$ 0.3$&$ 0.001$& $0.0005$& $\sim 10^{-4}$ &Ba12\\
  $35317$ &    $55.71$&   $2.39$ &     $17.95$&  $0.77$&  $6.08$& $0.51$ &       $15.0$& $0.1$ &    $\sim 10^{-4}$& $\sim 10^{-4}$ &$\sim 10^{-4}$  &Ba12\\
  $175740$ &   $81.97$ & $1.75$ &      $12.2$&  $0.26 $&        $2.95$& $0.26  $&     $ -9.18$ &$ 0.25$& $\sim 10^{-4}$& $\sim 10^{-4}$& $\sim 10^{-4}$ &Br10+Ba12 \\
  $199881$ &   $72.2$   &  $3.65$&      $13.85$&  $0.7  $&       $2.64$& $0.8   $&      $ -15.7$ &$ 0.3 $&$\sim 10^{-4}$&$\sim 10^{-4}$& $\sim 10^{-4}$ &Ba12\\
 $101197$ &    $82.99$ &  $6.82$ &      $12.05$&  $0.99 $&   $5.66$& $0.62 $&      $7.5$ &$ 0.3$& $\sim 10^{-4}$ & $\sim 10^{-4}$& $\sim 10^{-4}$  &Ba12\\
 $105678$ &    $74.02$ & $1.7$  &       $13.51$&  $0.31$&    $5.82$& $0.26 $&     $-17.4$ &$ 0.5 $&$\sim 10^{-4}$& $\sim 10^{-4}$&$\sim 10^{-4}$ & Ba12\\	
 $219828$ &    $72.31$ & $3.87$&     $13.83$&   $0.74  $&   $ 5.86$& $0.77$&     $  -24.14$ &$ 0.17$& $\sim 10^{-4}$&$\sim 10^{-4}$& $\sim 10^{-4}$ &Ba12\\
 $28676$ &    $38.7$ &    $0.88 $ &    $25.84$& $0.59$&  $4.47$& $0.73$&         $6.71$&$ 0.09$&$\sim 10^{-4}$&$\sim 10^{-4}$&$\sim 10^{-4}$  &Br10+Ba12\\
 $52242$  &    $68.17$ & $2.74$ &      $14.67$& $0.59$&   $5.07$&  $0.64$&       $31.3$ &$ 0.9$& $\sim 10^{-4}$& $\sim 10^{-4}$& $\sim 10^{-4}$ &Ba12\\
 $95915$ &     $66.62$ &  $2.13$ &       $15.01$&  $0.48$&   $5.09$& $0.53$&       $16.9$ &$ 0.3 $&$\sim 10^{-4}$& $\sim 10^{-4}$& $\sim 10^{-4}$  &Ba12\\
 $105000$ &    $71.07$  & $2.98$ &     $14.07$&  $0.59$&    $4.73$&   $0.75 $&    $ -14.8$ &$1.5$&$\sim 10^{-4}$& $\sim 10^{-4}$&$\sim 10^{-4}$  &Ba12\\
 $148317$ &    $79.62$ & $3.49$&       $12.56$&  $0.55 $&     $3.45$& $0.69$&      $-37.6$ &$ 0.4 $& $\sim 10^{-4}$& $\sim 10^{-4}$&$\sim 10^{-4}$  &Ba12\\
 $44821$ &     $29.33$& $0.53$  &     $34.1$& $0.62$&     $5.0 $&  $0.44 $&       $18.3$& $ 0.76$& $\sim 10^{-4}$& $\sim 10^{-4}$&$\sim 10^{-4}$  &Br10+Ba12\\	
 $68814$   & $ 80.45$ & $7.57$&  $12.43$ & $1.17$&  $3.65$ & $1.03$& $34.5$ & $0.3$ & $\sim 10^{-4}$& $\sim 10^{-4}$&$\sim 10^{-4}$ &Liu15\\
 $7735$ &      $85.69$ & $ 8.81$ &    $11.67$& $1.2$&    $3.5$& $1.18$ &        $21.7$ &$1.4$&$\sim 10^{-4}$& $\sim 10^{-4}$&$\sim 10^{-4}$   &Ba12 \\
 $100382$&   $93.98$&  $3.0$&           $10.64$& $0.34$&     $4.89$&  $0.35$&     $-10.9$&$0.4$& $\sim 10^{-4}$& $\sim 10^{-4}$& $\sim 10^{-4}$ &Br10\\
 $199951$ &    $70.22$& $1.28 $ &      $14.24$&  $0.26 $&      $1.78$& $0.21 $&      $ 17.6$ &$ 0.8  $&$\sim 10^{-4}$&$\sim 10^{-4}$& $\sim 10^{-4}$  &Ba12\\
 $168769 $ &  $50.18$ &  $3.7 $ &       $19.93$& $1.47 $&      $2.14$& $1.33 $&     $26.4$ &$ 0.2 $& $\sim 10^{-4}$& $\sim 10^{-4}$&$\sim 10^{-4}$  &Br10\\
 $46100$  &   $55.46$ & $2.61$ &    $18.03$& $0.85 $&    $9.35$&  $0.94$&       $21.3$ &$ 0.3 $& $\sim 10^{-4}$& $\sim 10^{-4}$& $\sim 10^{-4}$ &Ba12\\
  $83423 $ &    $72.1$ &  $4.94 $ &       $13.87$& $0.95$&   $7.96$& $1.2  $&       $-7.3$ &$ 3.4 $&$\sim 10^{-4}$& $\sim 10^{-4}$&$\sim 10^{-4}$  &Bo11+Ba12\\
  $91320 $  &   $90.5$ & $ 6.88$   &      $11.05$& $0.84 $&   $5.18$& $0.63$&       $17.5$ &$ 0.4$&$\sim 10^{-4}$& $\sim 10^{-4}$&$\sim 10^{-4}$  &Br10\\
   $102928$ &    $91.41$ &  $4.18 $ &     $10.94$&  $0.5$&     $0.63$&  $0.34 $&     $14.12$ &$ 0.06$&$\sim 10^{-4}$& $\sim 10^{-4}$&$\sim 10^{-4}$  &Br10\\
  $168442$ &   $19.56$ & $0.62$ &       $51.12$&  $1.63  $&   $2.3 $& $1.56 $&       $-13.8$ &$0.3 $&$\sim 10^{-4}$& $\sim 10^{-4}$&$\sim 10^{-4}$   &Br10 \\
  $154747$ &    $97.85$ &  $8.9$ &        $10.22$&  $0.93$&      $8.58$&   $0.78$&     $-14.9$& $0.3$&$\sim 10^{-4}$& $\sim 10^{-4}$& $\sim 10^{-4}$  &Ba12\\
  $183140$ &   $71.84$ & $6.61$&       $13.92$&  $1.28 $&     $13.97$& $0.91 $&   $ -28.8$ &$ 0.4 $& $\sim 10^{-4}$& $\sim 10^{-4}$&$\sim 10^{-4}$   &Ba12\\
    \\ \hline
  \end{tabular}\\
  \raggedright{$^1$ Br10= \cite{brown10}; \hspace{2mm} Bo11= \cite{bobylev11}; \hspace{2mm} Ba12= \cite{batista12}; \hspace{2mm} Liu14= \cite{liu15}}
  \end{minipage}
    \end{table*}

We now use the updated selection criterion from Eq.\ \ref{eq:SScriterion} to examine the stars that
have been proposed in the literature as solar sibling candidates.  The results are shown in table
\ref{tab:SS_candidates}. In the first column we list the names of the solar siblings candidates.
From the second to the ninth columns we show the value and uncertainty of their heliocentric
distances, parallaxes, proper motions and radial velocities respectively. These values were obtained
from the \textsc{simbad} catalogue \citep{simbad}. The tenth column lists mean value of \SF\ for
each star, given its coordinates in the space of $\varpi$, $\mu$ and $V\mathrm{r}$. The
corresponding RSE and the survival fraction for that region of phase space are shown in the eleventh
and twelfth columns respectively.

Note that the stars HD 147443 and HD 196676 have phase space coordinates corresponding to sibling
fractions of $0.76\pm 0.20$ and $0.56\pm0.38$, respectively. Their ages and metallicities are also
consistent with those of the Sun \citep{ramirez14}. However, given that these stars do not have
solar chemical composition \citep{ramirez14}, we can not identify them as solar siblings. This is
consistent with the fact that the value of \SF\ for these stars still allows for a significant
fraction of stars that are not solar siblings located in the same region of phase space.

Conversely, \cite{ramirez14} found that the stars HD 28676, HD 91320, HD 154747 and HD 162826
have the same age, metallicity and chemical composition as the Sun, within the observational errors.
However, according to the numbers in Table \ref{tab:SS_candidates} these stars have a low
probability of being solar siblings. This also holds for the star HD 68814, which is chemically
homogeneous with the Sun \citep{liu15} but is located in a phase space region where $\SF\sim
10^{-4}$. \textcolor{red}This discrepancy may be due to the limitations in our simulations, which may lead to
underestimates of \SF\ (see Sect. \ref{sect:application}) or may be attributed to the observation that there is chemical abundance
overlap between different clusters \citep{blanco15}, which implies the presence of stars that look
like solar siblings even if their phase space properties are very different.

From the small number of stars examined as potential solar siblings it is not possible to draw
further conclusions. For more progress on this issue the results of \gaia\ and the complementary
abundance surveys, such as \galah, will have to be awaited.

\subsection{Applicability of the sibling selection criteria}
\label{sect:application}

We have shown in this study that despite uncertainties in the Galactic potential parameters and
solar birth cluster initial conditions, it is possible to identify a region in the space of
parallaxes, proper motion, and radial velocities which is robustly predicted to contain a high
fraction of solar siblings with respect to disk stars. However, the selection criterion shown in
Eq.\ \ref{eq:SScriterion}  is only valid for the cluster initial conditions and Galaxy models
considered here. Changes in the mass and size of the Sun's birth cluster or in the modelling of the
Milky Way, might alter the region in phase-space where it is more likely to identify solar siblings.
For instance, massive clusters (with $10^4$ stars) evolving in the Galactic potential described in
Sect.\ \ref{sect:Gmodel} might have lifetimes of around $20$~Gyr \citep{gieles07}. Thus, after
$4.6$~Gyr of evolution, most of the solar siblings would still be bound to the cluster, showing a
clumped distribution in the phase-space for most of the Galactic parameters. Conversely, small open
clusters (as those described in Sect.\ \ref{sect:bcluster}) only survive a few Myr in a Galaxy
model containing transient spiral structure and giant molecular clouds \citep[see
e.g.][]{gieles06, lammers06, gieles07, kruijssen}. In such a more realistic potential the solar
siblings would be more dispersed in both radius and azimuth, completely mixed with other disk stars,
which would (much) lower the mean value of \SF\ in any given region of $(\varpi, \mu,
V_\mathrm{r})$. Another limitation is that we do not consider the vertical motion of the Sun and the vertical force of the bar and spiral arms in the cluster simulations. Although the solar siblings are
stars that move within the Galactic disk, the mean value of $f_\mathrm{sib}$
might change when considering a three-dimensional potential for
the Galaxy. For the types of solar birth clusters studied in this work the results thus strongly
support the need for chemical abundance surveys to attempt to identify the sun's siblings (and other
disrupted clusters).

One could consider making more sophisticated phase space searches for the solar siblings by making
use of conserved quantities (energy, angular momentum). However, if open clusters contribute a
significant fraction of the stars to the Galactic disk (and all stars existing on somewhat similar
orbits) it is not obvious that disrupted open clusters would stand out in integrals of motion
spaces. Our simple selection criterion also has the advantage of being defined entirely in the space
of observables where the properties of the errors are well understood.

\section{Summary}
\label{sect:concl}

We used numerical simulation to study the evolution and disruption of the Sun's birth cluster in the
Milky Way. In the simulations we include the gravitational force among the stars in the cluster and
the stellar evolution effects on the cluster population. We also include the external tidal field of
the Galaxy, which was modelled as an analytical potential containing a bar and spiral arms. We used
two Galactic models: one in which the Galaxy has two or four spiral arms and a ($2+2$) composite
model in which two spiral arms have smaller strength and pattern speed than the other two arms. The
aim of this study is to predict the present-day phase space distribution of the solar siblings (as
observed in astrometry and radial velocities) and to understand how \gaia\ data might be used to
pre-select solar siblings candidates for follow-up chemical abundance studies.

We found that the dissolution time-scale of the Sun's birth cluster is insensitive to the details of
the Galactic model, in particular to the parameters of the bar and spiral arms. For the set of
simulations carried out in this study, the Sun's birth cluster is completely disrupted in a
time-scale of $0.5-2.3$~Gyr, where the differences are due to different eccentricities and
perigalactica of the cluster orbits. 

After the dissolution of the Sun's birth cluster, the solar siblings move independently within the
potential of the Galaxy. Depending on the Galactic parameters, the solar siblings may currently be
more or less dispersed in Galactic radius and azimuth. If the orbits of the solar siblings are not
influenced by the \crsp\ or by the \olrbar, the present-day distribution of the solar siblings is
such that most of these stars are in the close vicinity of the Sun. Conversely, if the orbits of the
solar siblings are influenced by these two resonances, the current spatial distribution of the
siblings is more dispersed in radius and azimuth, with substructures in some regions of the Galactic
disk (this is also observed in the ($2+2$) composite model). In Galaxy models with four spiral arms,
the solar siblings are spread all over the Galactic disk.

We predicted the \gaia\ observations (astrometry and radial velocities) of solar siblings brighter
than $G=20$~mag. We use the GUMS simulation \citep{gums} to generate a large sample of stars which
mimic the disk stars that \gaia\ will observe. With this information, we computed the sibling
fraction $\SF= N_\mathrm{sib}/N_\mathrm{disk}$, which can be interpreted as the probability of
finding solar siblings in a certain region of the space of $\varpi$, $\mu$ and $V_\mathrm{r}$.
Regions in this phase-space where $\SF>0.5$ indicate that a large fraction of stars located there
might be solar siblings. Thus exploring those regions would increase the success rate in finding
solar siblings candidates in the future. We found that $\SF>0.5$ when $\varpi \geq 5$~mas, $4 \leq
\mu \leq 6$~masyr$^{-1}$, and $-2 \leq V_\mathrm{r} \leq 0$~km~s$^{-1}$. This result is very similar
to that by \cite{brown10} but is now obtained for a large fraction of simulations covering
a broad range of Galactic parameters and initial conditions for the Sun's birth cluster.

However, this selection criterion is only valid under the assumptions made in this study.
Introducing more realism into the simulations (transient spiral arms, molecular clouds) would lower
\SF\ and make the pre-selection of solar siblings on the basis of distance and kinematic data very
inefficient (unless the sun's birth cluster was originally much more massive). This reinforces the
conclusion already reached by \cite{BH10} that large scale surveys are needed which are aimed at
precisely determining the astrophysical properties of stars, in particular their ages and chemical
abundances, if we want to identify the solar family.

\section*{Acknowledgements}
We thank the anonymous referee for his/her suggestions that greatly improved the manuscript. This
work was supported by the Nederlandse Onderzoekschool voor Astronomie (NOVA), the Netherlands
Research Council NWO (grants \#639.073.803 [VICI], \#614.061.608 [AMUSE] and \#612.071.305 [LGM])
and by the Gaia Research for European Astronomy Training (GREAT-ITN) network Grant agreement no.:
264895.

\bibliographystyle{mn2e}
\bibliography{references}

\label{lastpage}
\bsp
\end{document}